\documentclass[times, twoside]{zHenriquesLab-StyleBioRxiv}

\usepackage[version=4]{mhchem} 
\usepackage{xspace}
\usepackage{algorithm}
\usepackage{algorithmicx}
\usepackage{algpseudocode}
\usepackage{newfloat}

\DeclareFloatingEnvironment[name=Extended Data Fig.]{extfigure}
\crefname{extfigure}{Extended Data Fig.}{Extended Data Figs.}

\newcommand{\crngen}{GenAI-Net\xspace}
\newcommand{\species}[1]{{\bf\ce{#1}}}
\newcommand{\defeq}{\triangleq}
\newcommand{\startercrn}{starter I/O CRN\xspace}
\newcommand{\template}{template\xspace}
\newcommand{\templates}{templates\xspace}
\newcommand{\net}{\mathcal N}
\let\emptyset\varnothing

\title{\crngen: A Generative AI Framework for Automated Biomolecular Network Design}
\shorttitle{\crngen}
\leadauthor{Filo} 
\author[1, a]{Maurice Filo}
\author[1, a]{Nicol\`o Rossi}
\author[2, a]{Zhou Fang}
\author[1, \Letter]{Mustafa Khammash}

\affil[1]{Department of Biosystems Science and Engineering, ETH Zürich, 4056 Basel, Switzerland}
\affil[2]{State Key Laboratory of Mathematical Sciences, Academy of
Mathematics and Systems Science, Chinese Academy of Sciences,
Beijing, 100190, China}
\affil[a]{These authors contributed equally}

\begin{document}
\maketitle

\begin{abstract}
Biomolecular networks underpin emerging technologies in synthetic biology—from robust biomanufacturing and metabolic engineering to smart therapeutics and cell-based diagnostics—and also provide a mechanistic language for understanding complex dynamics in natural and ecological systems. Yet designing chemical reaction networks (CRNs) that implement a desired dynamical function remains largely manual: while a proposed network can be checked by simulation, the reverse problem of discovering a network from a behavioral specification is difficult, requiring substantial human insight to navigate a vast space of topologies and kinetic parameters with nonlinear and possibly stochastic dynamics. Here we introduce \crngen, a generative AI framework that automates CRN design by coupling an agent that proposes reactions to simulation-based evaluation defined by a user-specified objective. \crngen efficiently produces novel, topologically diverse solutions across multiple design tasks, including dose responses, complex logic gates, classifiers, oscillators, and robust perfect adaptation in deterministic and stochastic settings (including noise reduction). By turning specifications into families of circuit candidates and reusable motifs, \crngen\ provides a general route to programmable biomolecular circuit design and accelerates the translation from desired function to implementable mechanisms.
\end {abstract}

\begin{keywords}
\noindent
chemical reaction networks | automated circuit design | generative AI | reinforcement learning
\end{keywords}

\begin{corrauthor}
mustafa.khammash\at bsse.ethz.ch
\end{corrauthor}


\section*{Introduction}
With modern advances in gene editing and DNA synthesis technologies \cite{jinek2012programmable,cong2013multiplex,kosuri2014large,hoose2023dna, chehelgerdi2024comprehensive} the rational engineering of living systems has become a practical reality \cite{elowitz2000synthetic,gardner2000construction,becskei2000engineering,aoki2019universal, porter2011chimeric,roybal2016precision}. 
This capability is poised to transform diverse areas---from precision therapeutics and cell-based diagnostics to sustainable biomanufacturing and new biomaterials---with major implications for human health and the environment \cite{khalil2010synthetic,voigt2020synthetic}. 
A central prerequisite is the ability to design biomolecular circuits and reaction networks that implement desired functions while remaining sufficiently compact and robust to operate in complex cellular contexts. 
Yet this design task remains difficult: biomolecular circuits are inherently nonlinear and often non-modular, and their function emerges from a complex interplay between reaction topology and kinetic parameters. As a result, achieving a desired behavior still typically demands substantial manual iteration and trial-and-error, reflecting the lack of a general, principled design approach.

To date, most biomolecular circuits have been designed by combining human intuition with principles from applied mathematics and classical systems and control theory. 
This approach enabled seminal synthetic circuits such as the repressilator \cite{elowitz2000synthetic}, the genetic toggle switch \cite{gardner2000construction}, and synthetic autoregulation \cite{becskei2000engineering}, which helped launch the field of synthetic biology. 
More recently, control-theoretic integral feedback has enabled the design of biomolecular circuits that achieve stringent forms of homeostasis \cite{aoki2019universal,agrawal2019vitro,briat2016antithetic,briat2016design,ni2009control,jones2022robust, anastassov2023cybergenetic, frei2022genetic, salzano2025vivo,chang2026engineering}. 
Proportional–Integral–Derivative (PID) control has also been translated into biomolecular circuit architectures to tune transient dynamics and improve performance beyond steady-state regulation \cite{filo2022hierarchy,chevalier2019design,alexis2022design,martinelli2025multicellular,paulino2019pid,whitby2021pid,modi2021noise}. Furthermore, alternative actuation and sensing mechanisms have been explored to achieve similar improvements in dynamical response and noise suppression \cite{samaniego2021ultrasensitive,kell2023noise, filo2023hidden,filo2024anti,hancock2022stabilization}.
 
Beyond regulatory feedback motifs, biomolecular networks have been proposed for intricate dynamical responses and biochemical information processing, including minimum realizations \cite{oishi2011biomolecular}, noise filters \cite{zechner2016molecular},
linear classifiers \cite{cherry2025supervised},
logic circuits \cite{seelig2006enzyme,moon2012genetic, malekpour2023wplogicnet}, and even artificial neural networks \cite{chen2024synthetic,cherry2018scaling,vasic2022programming,qian2011neural,okumura2022nonlinear,anderson2021reaction,fan2025automatic,moorman2019dynamical,samaniego2024neural}. 
However, achieving these richer behaviors may require complex networks, and translating such designs to \textit{in vivo} settings remains challenging—motivating automated approaches that can generate diverse candidate mechanisms and expose complexity–performance trade-offs.

\begin{figure*}[ht!]
\centering
\includegraphics[scale=1]{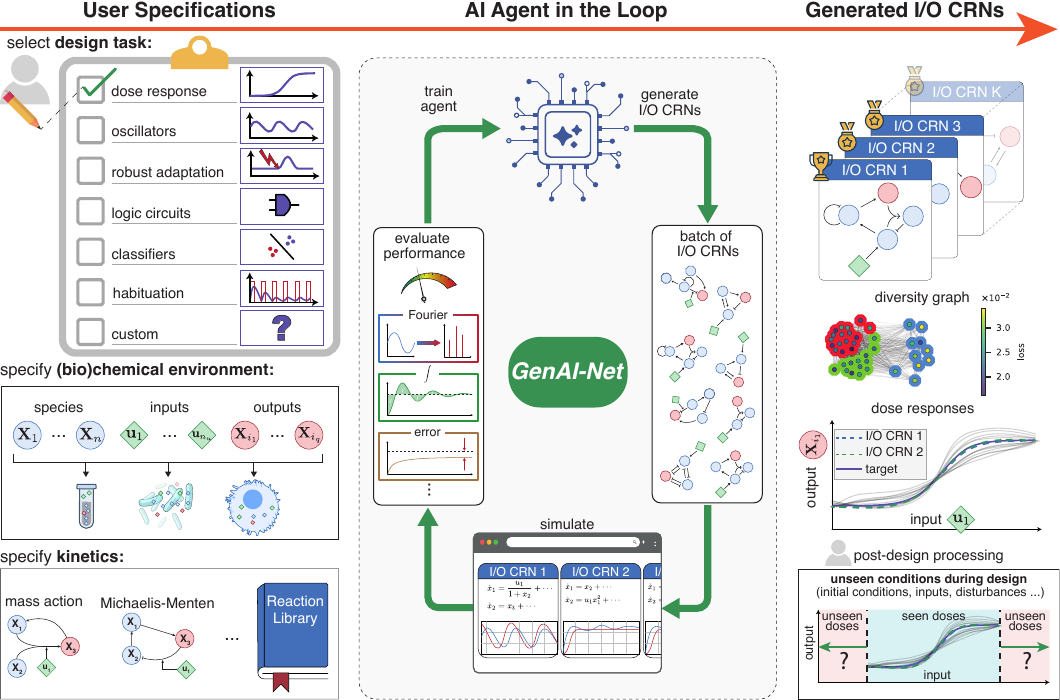}
\caption{\textbf{\crngen overview: a generative AI agent for automatic design of input--output chemical reaction networks (I/O CRNs).}
\textbf{User specifications.} Users begin by selecting a desired design \emph{task} (e.g. dose–response, oscillators, robust perfect adaptation, logic circuits, classifiers, habituation, or any custom task). They then specify the intended (bio)chemical operating environment, including the set of chemical species that may appear in the network, the designation of \emph{inputs} (e.g., externally set signals $u_1$) and \emph{outputs} (regulated/readout species), and any contextual constraints implied by the application setting (e.g. in vitro,  cellular contexts, or others). Finally, users choose an appropriate kinetic model class (e.g., mass-action or Michaelis–Menten) and provide (or select) an available \emph{reaction library} containing $M$ permissible reactions from which candidate networks may be assembled.
\textbf{AI agent in the loop.} Given these specifications, \crngen iteratively \emph{generates} candidate I/O CRNs, \emph{simulates} their dynamics under the chosen kinetics, and \emph{evaluates performance} against the task objective using quantitative metrics (e.g. time-domain error and frequency-domain/Fourier features). These evaluations provide a learning signal used to \emph{train the agent}, closing the design loop: the agent proposes new networks conditioned on prior performance, progressively improving the generated candidates. The output of a design run is a \emph{batch of I/O CRNs} sampled from the learned search policy, enabling downstream selection, analysis, and implementation.
\textbf{Generated I/O CRNs and behaviors.} \crngen returns multiple top-ranked candidate I/O CRNs, each represented by a specific reaction topology, parameters, and its corresponding predicted input--output behavior. Diversity graphs can used to monitor the topological diversity of the generated I/O CRNs. The example to the right highlights how topologically distinct candidate networks can realize similar target specifications (e.g., matching a desired dose–response curve) while differing in internal reaction structure, enabling users to trade off performance, simplicity, and implementability when choosing a final design. The resulting designs can then be evaluated under conditions not used in the design loop. For example, the bottom-right panel illustrates a post-design generalization test in which the generated I/O CRNs are simulated across dose levels that were not included during design.}
\label{Fig:Overview}
\end{figure*}

Beyond human intuition, biomolecular circuits can also be designed using machine intelligence. Early work \cite{ma2009defining} used exhaustive enumeration to identify adaptive behaviors across all three-node enzyme network topologies. As the number of species grows and enumeration becomes infeasible, learning-based strategies become essential. Other pipelines have leveraged sparse regression \cite{hiscock2019adapting}, evolutionary algorithms \cite{rossi2024synthevo, rodrigo2007genetdes, tatka2025speciated, paladugu2006silico}, tree search \cite{bhamidipati2025designing}, {topological filtering \cite{rybinski2020topofilter},} Thompson sampling \cite{kobiela2024risk}, and Bayesian optimization \cite{merzbacher2023bayesian} to discover biochemical networks with desired functions. More recently, a deep-learning approach based on conditional variational autoencoders was introduced \cite{gallup2025generative}, using simulations of three-node circuits to guide parameter search and retrieve circuits that exhibit adaptation \cite{gallup2025generative}. In a complementary direction, a supervised learning approach has been recently applied to large-scale experimental measurements to learn predictive composition-to-function models \cite{rai2026ultra}. Collectively, these methods demonstrate that machine intelligence can automate and accelerate biomolecular circuit design {\cite{palacios2025machine,rai2024using}}; however, existing approaches often remain constrained in either versatility (the range of tasks and network sizes they can handle) or efficiency (the ability to search large topology--parameter spaces without extensive task-specific engineering).

Often, the forward design direction is straightforward: given a proposed circuit, we can usually test whether it performs a desired task (e.g. by simulating its dynamics). The reverse direction—starting from a behavioral specification and discovering a network that realizes it—is far more challenging, because the space of reaction topologies and parameterizations is vast with nonlinear and possibly stochastic dynamics. Here we introduce \crngen, a generative AI framework that tackles this inverse problem by enabling flexible, automated design of chemical reaction networks (CRNs) for a broad range of user-defined tasks.
\crngen\ exploits the asymmetry in difficulty between the forward and reverse problems by placing an AI agent in the loop (Fig.~\ref{Fig:Overview}). The agent acts as an oracle to automatically propose candidate designs (reverse problem) and pass them to a simulator for performance evaluation (forward problem). The resulting evaluation signals are then used to refine the oracle’s accuracy and strengthen the agent’s generative capabilities.

A \crngen\ user starts by specifying a task-level objective—what behavior the CRN should realize—together with the chemical context in which the design must live (Fig.~\ref{Fig:Overview}, left). The user defines the molecular species and, when relevant, designated inputs and outputs, chooses a kinetic model (e.g., mass-action or Michaelis-Menten, deterministic or stochastic setting), and/or selects a reaction library that constrains which reaction types may be used. With these ingredients, \crngen\ automatically generates and returns a set of diverse candidate CRNs (Fig.~\ref{Fig:Overview}, right), enabling the user to compare alternative topologies and parameterizations and select designs that best match performance, simplicity, implementability, and even generalizability (Fig.~\ref{Fig:Overview}, bottom right).

We apply \crngen\ to a broad set of design specifications, including dose–response shaping (Hill-function matching, ultrasensitivity, and non-monotonic responses), robust perfect adaptation under setpoint changes and disturbances (both in deterministic models and in stochastic settings where noise reduction is additionally targeted), classifiers that map initial conditions to discrete fates, oscillators with fixed-mean dynamics or input-tunable frequency, logic circuits implementing complex Boolean gates, and habituation/sensitization. In each case, the specification is captured through a simulation-based performance evaluation (e.g., time-domain tracking error, classification loss over initial conditions, frequency-domain objectives, or truth-table accuracy), and \crngen\ successfully generates diverse CRN designs that meet the desired behaviors.
Together, these results position \crngen\ as a general-purpose engine for programmable molecular circuit design, enabling rapid discovery of diverse reaction motifs directly from high-level specifications. 
All the code to reproduce the results of this work, and to explore other applications is available in the following Github repository\footnote{\url{https://github.com/Maurice-Filo/GenAI-Net}} with an easy-to-use programmatic user interface and detailed documentation\footnote{\url{https://maurice-filo.github.io/GenAI-Net/}} and list of generated CRNs.

\section*{Results}

\begin{figure*}[ht!]
\centering
\includegraphics[scale=1]{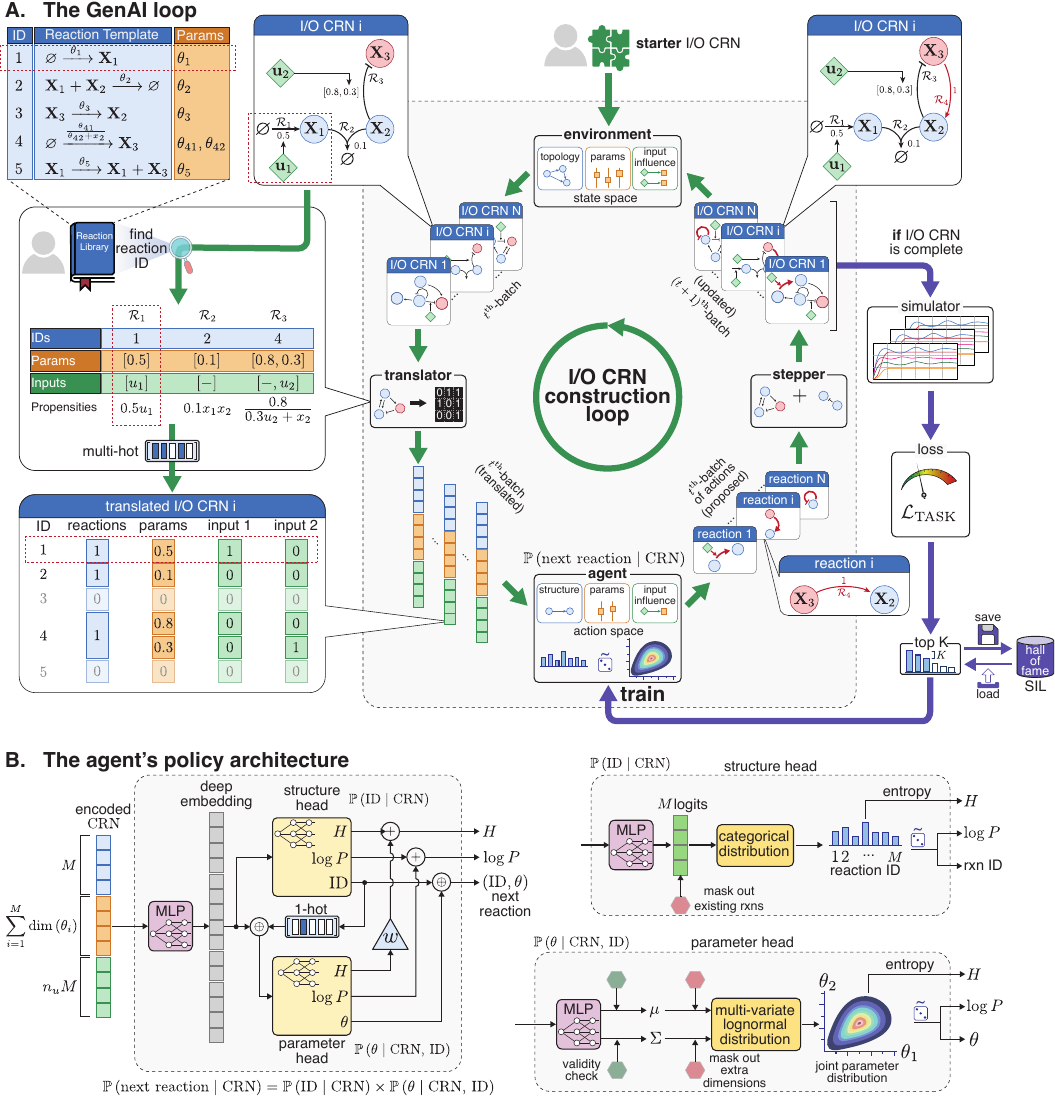}
\caption{\textbf{Method overview: the \crngen generative design loop and policy architecture.}
\textbf{(A) The GenAI loop.} Starting from a user-provided \startercrn (top), \crngen grows candidate networks by sequentially selecting reaction \templates from a \emph{reaction library} of size $M$ and assigning their kinetic parameters. A candidate I/O CRN (left; “I/O CRN $i$”) is represented as a set of reaction IDs and associated rate parameters (table), together with any designated external inputs that modulate propensities. This representation is translated into an agent-interpretable form (bottom left), where each reaction is encoded as a multi-hot entry indicating whether it is present, its parameter value(s), and which input channels influence its propensity (illustrated as “input 1” and “input 2”). This translation maps the I/O CRN state maintained by the environment into the fixed-format representation consumed by the agent. Conditioned on this translated representation, the \emph{agent} (bottom center) proposes an action (e.g., adding a new reaction and sampling its kinetic parameters). A \emph{stepper} (center right) applies the action to update the I/O CRN in the \emph{environment} (producing the next incomplete I/O CRN; top center), and the \emph{translator} (center left) is applied again to yield the agent’s next observation. Once a terminal condition is met (e.g. the required number of reactions to complete the I/O CRN is reached), the environment uses the updated I/O CRN to build the corresponding dynamical model under the chosen kinetic semantics and passes it to the \emph{simulator} (right)  and then to the loss module to evaluate the resulting trajectories and compute task-specific loss $\mathcal L_{\text{TASK}}$. Across many rollouts, \crngen maintains a ``hall of fame” of high-performing solutions, and uses these elite trajectories to refine the training of the agent via self-imitation learning (SIL), biasing future sampling toward successful reaction sequences and parameterizations seen in previous iterations.
\textbf{(B) The agent’s policy architecture.} The policy maps the current incomplete I/O CRN (encoded as the stacked multi-hot reaction/parameter/input tensor) to a \emph{deep embedding}, which feeds two coupled heads. The \emph{structure head} produces logits over the discrete action space of reaction IDs, defining a categorical distribution $\mathbb P(\mathrm{ID}\mid \mathrm{CRN})$; already-present reactions are masked to prevent redundant selection, and an entropy term encourages exploration during training. Given the selected reaction ID, the \emph{parameter head} outputs a joint distribution over the continuous kinetic parameters $\mathbb P(\boldsymbol{\theta}\mid \mathrm{CRN},\mathrm{ID})$. Together, these heads define the factored policy $\mathbb P(\text{next reaction}\mid \mathrm{CRN}) = \mathbb P(\mathrm{ID}\mid \mathrm{CRN}) \mathbb P(\boldsymbol{\theta}\mid \mathrm{CRN},\mathrm{ID})$, enabling end-to-end learning of both network topology and kinetics within the closed-loop generation–simulation–training pipeline.}
\label{Fig:Method}
\end{figure*}

In what follows, we demonstrate how \crngen\ enables automated, objective-driven discovery of chemical reaction networks across a wide range of dynamical tasks. We first present the core method—how candidate CRNs are represented, generated, and evaluated through simulation, and how this feedback is used to progressively improve design proposals. We then use this framework to produce diverse high-performing networks in multiple application settings, illustrating both the behaviors achieved and the reaction-level motifs that emerge from the search.

\begin{figure*}[ht!]
\centering
\includegraphics[scale=1]{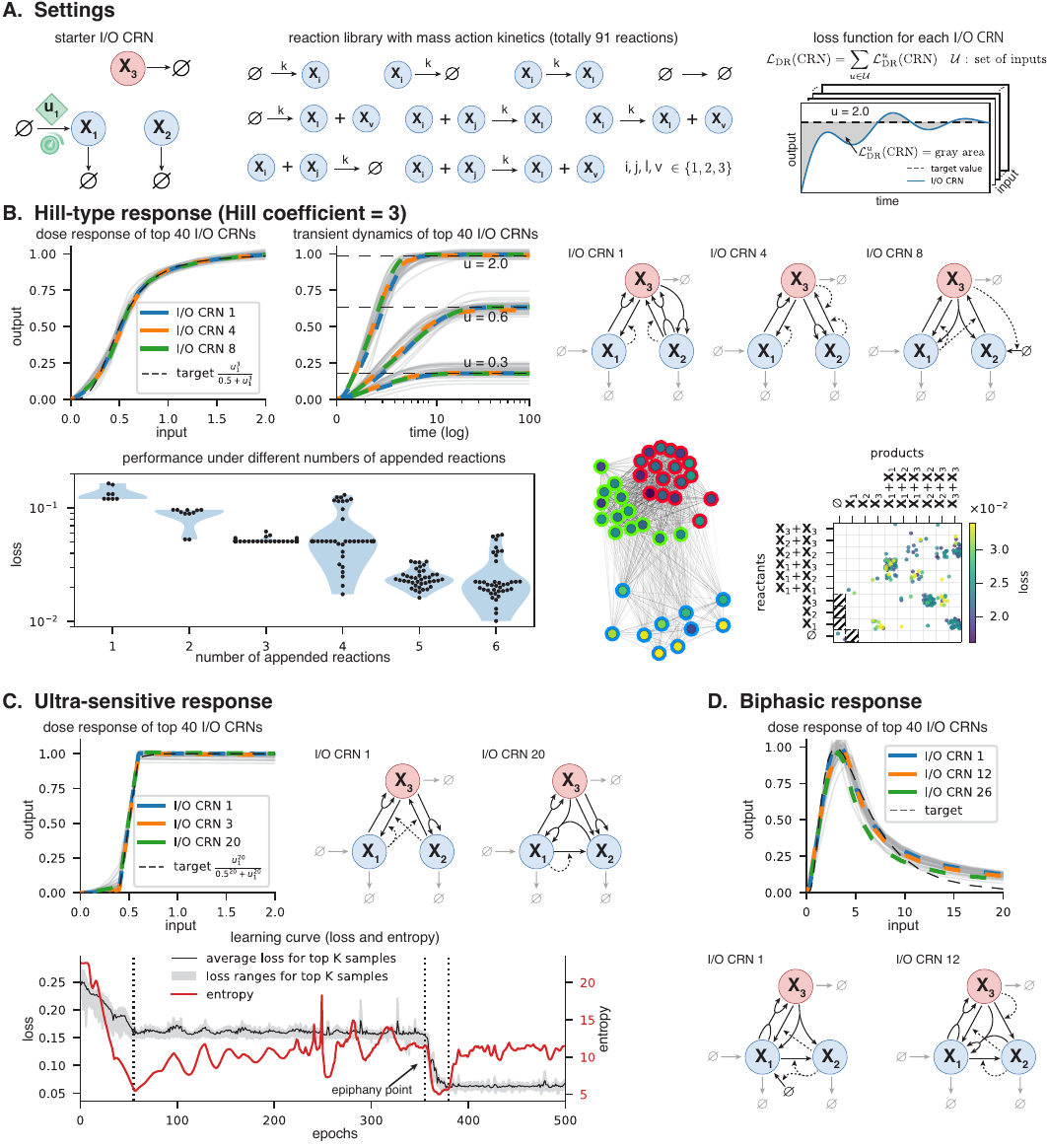}
\caption{\textbf{Dose Response.}
\textbf{(a) Problem setting for generating I/O CRNs exhibiting desired dose responses.}
\crngen\ starts from a \startercrn (left) consisting of three species, including an output species \species{X_3}, and four reactions. The rate of one of the reactions is modulated by an external input $u_1$.
From this \startercrn, \crngen\ generates candidate I/O CRNs by appending up to five doubly bimolecular mass-action reactions from a library (middle) and optimizes the average loss of the top 5\% generated networks together with an entropy term to boost network diversity.
The loss of each network is defined as a weighted $L_1$ norm of the difference between the system trajectory and the desired response over time, averaged across many different input conditions (see right). 
\textbf{(b) Generation of networks exhibiting Hill-type responses.}
\crngen\ is capable of generating more than 40 topologically unique networks whose steady-state dose-response closely matches the target Hill function (first panel in the first row), while also exhibiting fast and smooth transient dynamics (second panel in the first row). 
In these plots, the colored lines represent the performance of three relatively topologically distinct networks (top right), while the gray lines indicate the performance of the remaining networks among the top 40.
The graph visualization (the middle panel in the second row) summarizes topological diversity across the top 40 generated solutions: nodes denote I/O CRN topologies, edge grayscale encodes pairwise Hamming distance (the number of differing reactions), node fill color indicates loss, and node outline color indicates clusters identified automatically via the Louvain method.
The reactant--product incidence map (right bottom) summarizes reaction usage across the generated collection of I/O CRNs: each point corresponds to the reactants and their products, and points are colored by the loss associated with I/O CRNs in which the reaction appears (color bar; scale noted).
The bottom-left panel illustrates how the performance of the top I/O CRNs changes with the number of appended reactions.
\textbf{(c)--(d) Generation of networks exhibiting ultrasensitive and biphasic responses, respectively.}
The bottom panel in (c) shows the evolution of the loss and entropy over epochs during the learning procedure.
}
\label{Fig:dose_response}
\end{figure*}

\subsection*{Input-Output Chemical Reaction Networks (I/O CRN)}
We represent each design as an \emph{input–output chemical reaction network} (I/O CRN): a collection of interacting species connected by reactions, where some quantities are treated as external inputs that modulate reaction rates and one or more species are designated as outputs of interest (to be measured or regulated). Each reaction is defined by its stoichiometry and kinetics. 
More precisely, a single reaction $\mathcal R_j$ over species (\species{X_1}, $\ldots$, \species{X_n}) is written as
\begin{equation*}
\mathcal R_j: ~
\text{
\ce{
$\nu_{1j}$ \species{X_1} + $\cdots$ + $\nu_{nj}$ \species{X_n} ->[$\lambda_j(x,u;\theta)$] $\nu'_{1j}$ \species{X_1} + $\cdots$ + $\nu'_{nj}$ \species{X_n}}}, 
\end{equation*}
where $\nu$'s $\in\mathbb{N}_0$ are stoichiometric coefficients and $\lambda_j(x,u;\theta)$ is the reaction’s \emph{propensity} (rate), which depends on the current state $x$ (species concentration or molecular counts), optional external inputs $u$ that modulate reaction rates, and kinetic parameters $\theta$. Each time this reaction fires, the system state changes by the net stoichiometric update $\Delta_j=\nu'_j -\nu_j$. For a network with $m$ reactions, stacking these $\Delta_j$ vectors as columns yields the stoichiometric matrix $S\in\mathbb{Z}^{n\times m}$, and stacking the propensities yields $\lambda (x,u;\theta)\in\mathbb{R}^m_{\ge 0}$. In the deterministic setting, the resulting dynamics are $\dot{x}=S \lambda(x,u;\theta)$; in the stochastic setting, the same $S$ and $\lambda$ define a continuous-time Markov jump process described by the chemical master equation (CME), typically simulated with the Gillespie algorithm \cite{gillespie1977exact}. An I/O CRN is simply such a collection of reactions, together with a designation of which signals serve as inputs $u$ and which species are treated as outputs. 


\subsection*{The Reinforcement Learning (RL) Loop}
The RL loop (Fig.~\ref{Fig:Method}A) begins when the user specifies (i) a \startercrn and (ii) the maximum number of reactions $m$ that may be appended to complete the design. Together, these choices define the start of the construction process (the \startercrn) and its endpoint (a completed I/O CRN obtained after $m$ reaction additions).

At the core of \crngen\ is an \emph{environment} (top center of the construction loop in \Cref{Fig:Method}A) whose \emph{state} is the current I/O CRN, which may be incomplete. Initially, the state is exactly the user-provided \startercrn. As the loop unfolds, the state evolves through a discrete, combinatorial \emph{state space} consisting of all CRNs obtainable by appending up to $m$ reactions from the chosen reaction library, together with their associated kinetic parameterizations and any permitted input-modulation assignments. Thus, each state encodes the reactions selected so far (and how they are parameterized) while preserving the \startercrn's species and any designated input/output roles.

Because a CRN state is naturally expressed as a human-readable list of reactions, \crngen\ includes a \emph{translator} (center left of the construction loop in \Cref{Fig:Method}A) that converts the current I/O CRN into a fixed-format, agent-interpretable representation. Concretely, the translator maps the current set of selected reactions into a structured encoding that, for each library reaction, records whether it is present, its kinetic parameter value(s), and which input channels (if any) modulate its propensity. This translation standardizes variable-sized CRN descriptions into a consistent representation that the policy can condition on, while maintaining a direct correspondence back to reaction-level semantics.

Given the translated state, the \emph{agent} (bottom center of the construction loop in \Cref{Fig:Method}A) selects the next reaction to add. The \emph{action space} is defined by the reaction library: each action corresponds to choosing a reaction identity (and, when relevant, its parameter values and input modulation choices), subject to constraints such as masking reactions already present. The agent's decision is governed by a learned \emph{policy}. Importantly, although the agent consumes the translated encoding, it produces its decision in a \emph{human-interpretable} form---for example, in Fig.~\ref{Fig:Method}A (under proposed reaction) the agent selects the highlighted reaction (shown in red) to append next.

A lightweight \emph{stepper} block (center right of the construction loop in \Cref{Fig:Method}A) then applies this action by appending the proposed reaction to the existing CRN, producing the next environment state; in the running example, the stepper augments the template by adding the red reaction to yield the updated I/O CRN shown in Fig.~\ref{Fig:Method}A (under updated I/O CRN). This construction loop repeats for $m$ steps (or until a termination condition is met), yielding a \emph{complete} I/O CRN.
Once the I/O CRN is complete, \crngen\ switches from the outer construction loop to an inner \emph{evaluation-and-training} loop. The completed CRN is passed to a \emph{simulator}, its trajectories are scored by the task objective to produce a scalar \emph{loss}, and this loss provides the learning signal used to update the policy network of the agent (Fig.~\ref{Fig:Method}B).
For a more detailed description of the various components in the RL loop, see Methods.


\subsection*{\crngen generates networks with target dose-responses}
We first exemplify \crngen\ for generating networks with target dose-responses. 
This task is prevalent in biological studies, aiming to identify the underlying biochemical mechanisms given experimentally measured dose-response curves or generate networks exhibiting such responses for synthetic-biology applications. 
This problem is often hindered by the vast search space over both network structure and kinetic parameters, making exhaustive search impractical.  
For instance, we consider a system consisting of three chemical species and four starting reactions as shown in \Cref{Fig:dose_response}A, where the input signal regulates the production rate of one species, and the remaining degradation reactions (one for each reaction) represent the dilution process due to cell growth.
Allowing up to five more doubly bimolecular mass-action reactions yields more than $3\times 10^7$ possible networks (i.e., the number of combinations of 87 candidate reactions taken 5 at a time), and this space expands further when accounting for choices of reaction rate constants.

Despite these challenges, \crngen\ can generate a diverse set of networks that realize target dose–response curves, including Hill-type, ultra-sensitive, and non-monotonic behaviors (see \Cref{Fig:dose_response}B--D). As shown, for each target curve, \crngen\ identifies more than 40 topologically distinct networks whose responses closely match the desired profile.
This diversity and precision arise from the balance of the loss and entropy terms of the network in the objective function. 
In one representative training procedure (bottom panel of \Cref{Fig:dose_response}C) 
the entropy term (red) drops quickly when the loss (black) decays rapidly—most notably during the first and third periods separated by dashed lines—indicating strong exploitation to select high-performing I/O CRNs from previously sampled networks. In other periods, when the loss plateaus, the entropy gradually increases, signaling exploration for new topologies with lower loss and thereby yielding diverse, high-quality networks.
In addition to capturing the target dose response, the generated networks exhibit smooth, fast-converging transient dynamics (second panel in the first row of \Cref{Fig:dose_response}B). This is encouraged by the loss-function design (see Methods), which evaluates trajectory-level mismatches between simulated dynamics and target values over time (right panel of \Cref{Fig:dose_response}A).

The generated networks also exhibit recurring topological patterns, highlighting specific design principles. In particular, several reactions are frequently selected by high-performing networks; whereas others are consistently avoided (see the bottom-right panel of \Cref{Fig:dose_response}B for the Hill-type response). By grouping networks according to topological similarity using the Louvain method (bottom-middle panel of \Cref{Fig:dose_response}B), we identify several representative topologies across the resulting clusters, as shown in the top-right panel of \Cref{Fig:dose_response}B. Similar patterns arise in the other dose-response examples, with representative networks shown 
in \Cref{Fig:dose_response}C--D.

The number of appended reactions is a key design knob for navigating the trade-off between topological simplicity and achievable performance.
Allowing more appended reactions expands the space of candidate mechanisms and can improve the performance of the top I/O CRNs (bottom-left panel of \Cref{Fig:dose_response}B), with the best-performing network improving steadily as the reaction budget increases. This flexibility enables users to select a reaction budget that matches the desired balance between compactness and functional precision, depending on the requirements of the application.

\subsection*{\crngen\ discovers robust perfect adaptation circuits for setpoint tracking and disturbance rejection}

\begin{figure*}[ht!]
\centering
\includegraphics[scale=0.95]{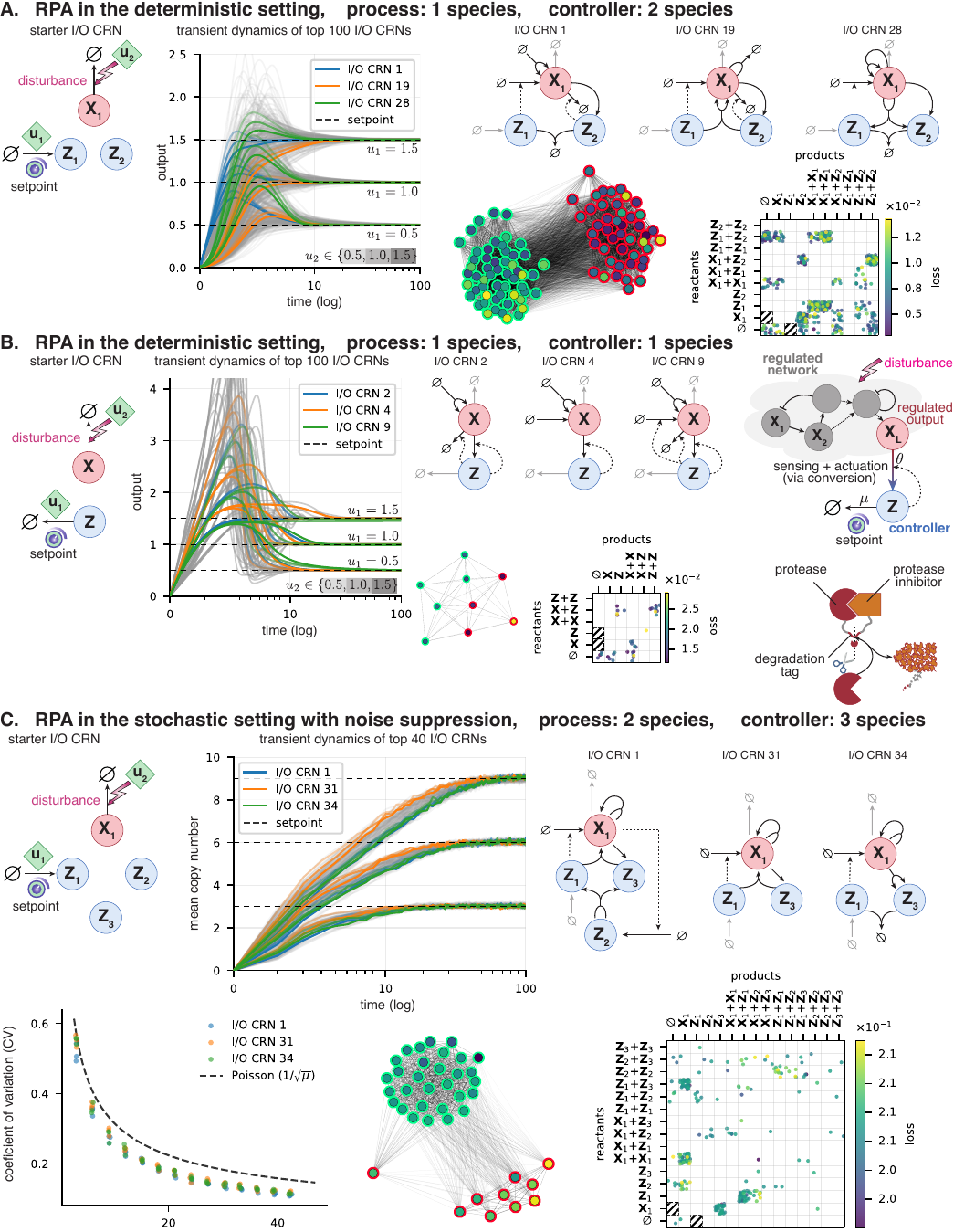}
\caption{\textbf{Robust perfect adaptation: setpoint tracking and disturbance rejection.} The Figure caption is on the next page.}
\label{Fig:RPA}
\end{figure*}

\begin{figure*}[ht!]\ContinuedFloat
\caption{\textbf{Robust perfect adaptation: setpoint tracking and disturbance rejection.}
\textbf{(a)} \crngen\ begins with a \startercrn (left) consisting of three species---the regulated output \species{X_1} (red) and controller species \species{Z_1}, \species{Z_2} (blue)---and two reactions whose rates are modulated by external inputs $u_1$ and $u_2$. The input $u_1$ tunes the desired setpoint by adjusting the production of \species{Z_1}, while $u_2$ acts as a disturbance by modulating the degradation rate of the output species \species{X_1}. From this \startercrn, \crngen\ generates candidate I/O CRNs by appending up to five doubly bimolecular mass-action reactions, optimizing a loss defined as a weighted $L_1$ norm of the setpoint-tracking error over time, averaged across three setpoints ($u_1 \in \{0.5,1.0,1.5\}$) and three disturbance magnitudes ($u_2 \in \{0.5,1.0,1.5\}$); see the loss definition in Methods. The time-course plot shows the regulated output versus log time for the top 100 topologically unique I/O CRNs (thin gray trajectories), with three representative solutions highlighted (colored curves; I/O CRNs 1, 19, and 28). For each highlighted I/O CRN, transparency encodes the disturbance strength set by $u_2$ (lighter to darker shades). All responses demonstrate robust convergence toward the desired setpoint levels (horizontal dashed lines) across the scanned disturbance and setpoint settings.
Example I/O CRN topologies for the highlighted networks are shown to the right. The graph visualization (bottom center) summarizes topological diversity across the 100 generated solutions. The reactant--product incidence map (bottom right) summarizes reaction usage across the generated collection of I/O CRNs.
\textbf{(b)} A minimal template with a single controller species \species{Z} yields a smaller \crngen-generated \emph{solution set} (9 topologically unique I/O CRNs). Here, up to four reactions are added. As in (a), the center panel shows output trajectories for all generated solutions (gray), with three representative networks highlighted (I/O CRNs 2, 4, and 9) and their corresponding topologies shown to its right. The reactant--product incidence map and the topological diversity graph are also shown. The schematic to the far right highlights a control motif discovered by \crngen (I/O CRN 4) that closely resembles the autocatalytic integral controller reported in \cite{briat2016design, drengstig2012robust}, yet departs from the standard control-theoretic separation between \emph{sensing} and \emph{actuation}. Here, a single molecular conversion reaction performs both operations simultaneously, effectively encapsulating the feedback computation within one reaction channel. The cartoon genetic implementation (bottom far right) illustrates one possible realization via protease-mediated degradation, in which a protease (and its inhibitor) targets a degradation tag to implement the required effective controller reactions.
\textbf{(c) Stochastic robust perfect adaptation with noise suppression: setpoint tracking with disturbance rejection while minimizing coefficient of variation.}
\crngen\ starts from a stochastic \startercrn (top left) consisting of an output species \species{X_1} (red) and three controller species \species{Z_1}--\species{Z_3} (blue). The input $u_1$ specifies the desired mean setpoint for \species{X_1}, while $u_2$ acts as a disturbance that must be rejected, i.e., changes in $u_2$ should not alter the steady-state mean of \species{X_1}. From this template, \crngen\ generates candidate I/O CRNs by appending additional reactions and optimizing a stochastic objective that enforces robust perfect adaptation (RPA) in the mean---setpoint tracking with rejection of $u_2$---while simultaneously minimizing output noise, quantified via the coefficient of variation (CV) of \species{X_1} (see Methods). The time-course panel (top center) shows mean trajectories of \species{X_1} versus log time for the generated \emph{solution set} (thin gray trajectories), with three representative solutions highlighted (colored curves; I/O CRNs 1, 31, and 34) and target setpoint levels indicated by horizontal dashed lines. Across scanned conditions, highlighted solutions robustly converge to the prescribed mean setpoints while maintaining adaptation despite the disturbance $u_2$. Noise performance is summarized in the CV panel (second row, left): for the three representative networks, the CV remains below the Poissonian reference level (black curve), indicating sub-Poisson fluctuations and improved noise suppression relative to the open-loop configuration where no controller species are present.
}
\noindent\rule{\linewidth}{0.4pt}
\end{figure*}

One of the fundamental tasks of synthetic biomolecular feedback controllers \cite{filo2023biomolecular} is to maintain homeostasis, a property with broad impact in bioproduction, metabolic engineering, and cell-based therapies, where many diseases arise from homeostatic failure \cite{kotas2015homeostasis, o2018homeostasis}. Robust perfect adaptation (RPA) \cite{khammash2021perfect, frei2021adaptive, xiao2018robust} is a stringent form of homeostasis: it enforces exact steady-state regulation of a target variable to a prescribed setpoint despite persistent disturbances. A substantial body of work has developed control-theoretic CRN designs that achieve RPA or near-RPA \cite{briat2016antithetic, briat2016design, ni2009control, drengstig2012robust, samaniego2021ultrasensitive, qian2018realizing,steel2019low,kelly2018synthetic,m2025multi,zand2024cascaded}, demonstrated experimental implementations \cite{aoki2019universal, huang2018quasi, anastassov2023cybergenetic, frei2022genetic, mallozzi2024crisprator, agrawal2019vitro, zhang2024crispr, chang2026engineering}, and provided increasingly general characterizations of RPA-capable network classes \cite{aoki2019universal, gupta2022universal, gupta2023internal, hirono2025rethinking, araujo2018topological}. However, even with these characterizations, the remaining design space can still be large, and satisfying structural conditions alone does not guarantee favorable transient performance.

In the following examples, we use \crngen\ to design molecular controllers that achieve RPA with good dynamic performance (Fig.~\ref{Fig:RPA}). We formulate RPA as an input--output control task in which an external input $u_1$ specifies the desired setpoint and a second input $u_2$ acts as a persistent disturbance by modulating the effective degradation of the regulated species (Fig.~\ref{Fig:RPA}A,B, left). With a minimal \startercrn containing the regulated output species and controller species, \crngen\ appends a user-defined number of additional reactions from the library and optimizes a trajectory-level loss (see Methods) that penalizes setpoint-tracking error over time, averaged across multiple setpoints and disturbance strengths. This objective explicitly rewards both fast and smooth dynamics and accurate steady-state regulation across conditions.

\paragraph{Single-process-species yield diverse RPA solutions.}
In Fig.~\ref{Fig:RPA}A, the template contains one process species \species{X_1} (regulated output) and two controller species \species{Z_1}, \species{Z_2}. Allowing up to five appended reactions, \crngen\ generates a large solution set: the top 100 topologically unique I/O CRNs (gray) robustly converge to the correct setpoints (dashed lines) across scanned $(u_1,u_2)$ conditions, with three representative networks (I/O CRNs 1, 19, and 28) highlighting distinct topologies that realize the same macroscopic behavior. The diversity graph and reactant--product incidence map summarize structural relationships among solutions and recurrent reaction usage in low-loss designs. Notably, the frequently selected sequestration reaction \ce{\species{Z_1} + \species{Z_2} -> $\emptyset$} effectively rediscovers the antithetic integral feedback (AIF) motif \cite{briat2016antithetic}. At the same time, \crngen\ repeatedly selects alternative sensing and actuation mechanisms that improve transients, including a conversion-based sensing reaction \ce{\species{X_1} -> \species{Z_2}} (rather than the more common catalytic sensing of \species{X_1}) and modified ``sequestration'' variants in which \species{Z_1} and \species{Z_2} jointly contribute to actuation of \species{X_1} instead of purely annihilating. Together, these results illustrate how \crngen\ can simultaneously satisfy stringent steady-state objectives (RPA) while discovering reaction-level modifications that yield favorable dynamics. 

\paragraph{A minimal single-controller \startercrn reveals compact motifs.}
Fig.~\ref{Fig:RPA}B reduces the \startercrn\ to a single controller species \species{Z}, yielding a smaller but still nontrivial solution set (9 topologically unique I/O CRNs). Despite this minimal starting point, \crngen\ still discovers networks that robustly adapt across setpoints and disturbances, with representative solutions shown (I/O CRNs 2, 4, and 9). Notably, one topology (I/O CRN 4) realizes a particularly compact feedback motif: a \emph{single molecular conversion reaction} simultaneously (i) senses the regulated output and (ii) drives the compensatory action, collapsing sensing and actuation into one reaction channel. This motif is closely related in function to autocatalytic integral control \cite{briat2016design, drengstig2012robust}, but differs in its \emph{non-modular} implementation: rather than a ``sense--compute--actuate'' separation, the feedback computation is inseparable from the actuation biochemistry itself. The schematic in Fig.~\ref{Fig:RPA}B (right) illustrates this combined conversion-based mechanism, together with one possible protease-mediated genetic realization in which a protease and its inhibitor form a complex whose conversion back to active protease implements integral-like accumulation while maintaining robustness to persistent disturbances.

\paragraph{Scaling to higher-dimensional regulation.}
Finally, Extended Data Fig.~\ref{EDF:RPA} demonstrates that \crngen\ scales to a multi-species setting with two process species \species{X_1}, \species{X_2} (with \species{X_2} as the regulated output) and four controller species (\species{Z_1}$,\dots,$ \species{Z_4}), where multiple disturbances modulate degradation of each process species. Starting from this higher-dimensional \startercrn and allowing eight additional appended reactions, \crngen\ again generates a large set of topologically distinct solutions (top 100 unique I/O CRNs) that achieve robust setpoint tracking for the regulated output across the scanned disturbance conditions, with representative topologies shown alongside the aggregate diversity summaries.

Together, these RPA examples show how \crngen\ can (i) generate \emph{families} of mechanistically distinct adaptive and high performance controllers from a common \startercrn, (ii) recover compact control motifs from minimal starting points, and (iii) scale the same design principles to higher-dimensional regulation problems. We also successfully tested other adaptation tasks, namely habituation and sensitization \cite{smart2024minimal, eckert2024biochemically} where inputs are allowed to be time varying. The results are reported in the Methods section and in Extended Data Fig.~\ref{EDF:Habituation}.

\subsection*{\crngen\ generates RPA networks with reduced coefficient of variation in the stochastic reaction kinetics}

Intracellular reaction systems can exhibit strongly stochastic dynamics when molecular copy numbers are low \cite{mcadams1997stochastic}, in contrast to the deterministic regimes considered in our previous examples. Such intrinsic noise can be fundamentally difficult to suppress \cite{lestas2010fundamental}.
To evaluate \crngen\ under these conditions, we tasked the method with constructing RPA networks with tunable means and reduced \textit{Coefficients of Variation} (CV) over time, a property often desirable experimentally as it mitigates dynamical intrinsic noise that manifests as cell-to-cell variability.
Similar to the RPA task in the deterministic setting, we selected a \startercrn with one output species, three controller species, and two starting reactions regulated by two inputs representing the setpoint and disturbance, respectively (\Cref{Fig:RPA}C top-left).
Allowing up to six appended reactions, \crngen\ discovers more than 40 topologically distinct controllers that achieve stochastic RPA: the mean trajectory tracks the setpoint across disturbance levels $u_2$ (Fig.~\ref{Fig:RPA}C, top middle). At the same time, the corresponding coefficients of variation (CVs) remain consistently below the Poisson limit $1/\sqrt{\text{mean}}$, i.e., below the open-loop noise level for \species{X_1} in the absence of controller species. This demonstrates that \crngen\ can jointly satisfy stringent steady-state regulation (RPA), maintain favorable transient performance, and attenuate intrinsic noise under disturbances.

\begin{figure*}[ht!]
\centering
\includegraphics[scale=1]{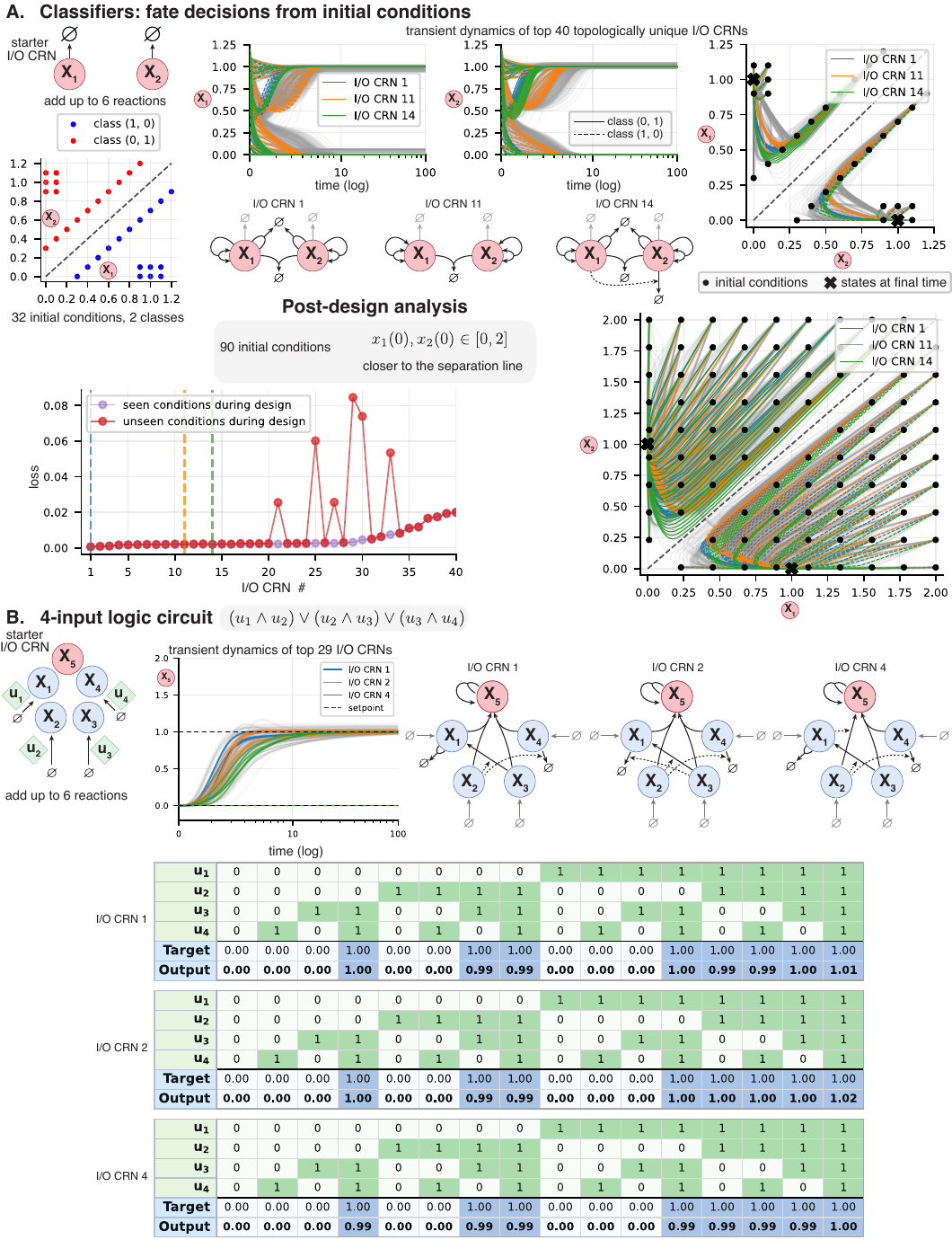}
\caption{\textbf{Classifiers and logic circuits.} The Figure caption is on the next page.}
\end{figure*}

\begin{figure*}\ContinuedFloat
\centering
\caption{\textbf{Classifiers and logic circuits.}
\textbf{(a) Molecular classifiers from initial-condition--dependent fate decisions.}
\crngen\ is tasked with generating I/O CRNs that \emph{classify} the initial state of a two-species system into one of two discrete outcomes. The \startercrn (top left) contains two output species, \species{X_1} and \species{X_2} (red), and no external inputs, so classification is driven solely by the network’s intrinsic nonlinear dynamics. A set of 32 initial conditions in the \species{X_1}--\species{X_2} plane (left) is partitioned into two classes (blue: class $(1,0)$; red: class $(0,1)$) corresponding to desired long-time fates in which one species is high while the other is low (diagonal decision boundary shown). From the \startercrn, \crngen generates candidate I/O CRNs by appending up to 6 doubly bimolecular reactions with mass-action kinetics, and optimizing a task loss that penalizes incorrect assignment of initial conditions to the target fate basins (see Methods for the the loss function).
The time-course plots (top center) show \species{X_1} (left) and \species{X_2} (right) trajectories versus log time for the top 40 topologically unique generated I/O CRNs (thin gray curves), with three representative solutions highlighted (colored; I/O CRNs 1, 11, and 14). For each highlighted I/O CRN, trajectories are shown across the full set of labeled initial conditions, demonstrating separation into two attractors consistent with the target classes (solid versus dashed curves). The phase-plane visualization (top right) summarizes the same simulations in the \species{X_1}--\species{X_2} plane, where points indicate trajectories from the labeled initial conditions and the two terminal regions correspond to the two classes (crosses); example reaction topologies for the highlighted I/O CRNs are shown in the center.
\emph{Post-design analysis} (bottom): to test generalization beyond the initial conditions seen during the design process, the generated I/O CRNs are evaluated on an expanded set of 90 initial conditions, some of which are concentrated near the decision boundary. The loss-versus-network index plot contrasts initial conditions seen during design (purple) with unseen initial conditions (red), and the phase-plane plot shows the resulting terminal states, highlighting how the learned basin structure extrapolates to nearby initial conditions not seen during the design.
\textbf{(b) Four-input irreducible logic circuit discovery.} \crngen\ begins with a \startercrn (top left) in which four external inputs $u_1$--$u_4$ act on corresponding input species \species{X_1}--\species{X_4} (blue) to regulate an output species (red). From this \startercrn, \crngen\ generates candidate I/O CRNs by appending additional reactions and optimizing a loss (see Methods) that rewards correct realization of the target 4-input irreducible Boolean function $(u_1 \wedge u_2)\ \vee\ (u_2 \wedge u_3)\ \vee\ (u_3 \wedge u_4)$. The time-course panel shows the output response versus time for the discovered solution set (thin gray trajectories), with three representative solutions highlighted (colored curves) and the low/high digital setpoints indicated by horizontal dashed lines. Example I/O CRN topologies for the highlighted networks are shown to the right, together with their truth tables (bottom), demonstrating high-precision binarization of the output across all 16 input combinations.}
\label{figure:Classifier}
\noindent\rule{\linewidth}{0.4pt}
\end{figure*}

The resulting high-performing networks exhibit recurring motifs reminiscent of antithetic integral feedback \cite{briat2016antithetic}, yet go beyond it in a direction that explicitly targets variance reduction. Whereas classical antithetic integral feedback is known to increase noise relative to open loop \cite{briat2016antithetic}, prior work has sought to mitigate this through added circuitry (e.g., molecular PID) \cite{filo2022hierarchy, briat2018antithetic}, modified actuation \cite{kell2023noise, filo2023hidden}, or auxiliary antithetic controllers that regulate higher moments \cite{lim2025toward}. In contrast, \crngen\ automatically uncovers motifs that simultaneously address all three objectives—RPA, fast dynamics, and noise suppression. As in other tasks, we cluster the top-performing solutions by topological similarity (Fig.~\ref{Fig:RPA}C, bottom middle) and report representative circuits (Fig.~\ref{Fig:RPA}C, top right).

Mechanistically, these circuits repeatedly employ conversion-based sensing \ce{\species{X_1} -> \species{Z_2}} (rather than the more common catalytic sensing), and in some families replace pure annihilation \ce{\species{Z_1} + \species{Z_2} -> $\emptyset$} with a coupled comparison--actuation step \ce{\species{Z_1} + \species{Z_2} -> \species{X_1}}, which both removes controller species and directly actuates \species{X_1}. In addition, the networks consistently incorporate negative autoregulation on \species{X_1} (Fig.~\ref{Fig:RPA}C, top right and bottom right), providing proportional negative feedback that suppresses stochastic fluctuations. Other solutions add parallel sensing routes—for example, an intermediate controller species that catalytically senses the output while producing both controller species—yielding alternative implementations that preserve adaptation while improving transient and noise properties.

\subsection*{\crngen generates circuits for fate decision}
\crngen\ can also generate biochemical circuits that implement initial-condition-dependent fate decisions, a key mechanism in cell differentiation and tissue development \cite{bier2015bmp}. 
Substantial effort has been devoted to the theoretical design and experimental implementation of biomolecular circuits that perform decision making through a range of dynamical mechanisms, including multistable behavior \cite{otero2017chemical, kim2006construction, santos2020multistable, wu2013engineering, oyarzun2015design, siegal2009capacity}.
To illustrate this capability, we tasked \crngen\ with designing a system of two species and a \startercrn compised of two reactions (\Cref{figure:Classifier}, top left).
The resulting network is required to converge to one of two fixed points depending on the initial conditions: if the first species initially dominates, the system converges to $(1,0)$; otherwise, it converges to $(0,1)$.
Using this setup and a loss function (see Methods) that evaluates performance from several selected initial conditions (\Cref{figure:Classifier} middle left), \crngen\ can generate more than 40 topologically unique, high-performing networks whose dynamics satisfy the above requirement (\Cref{figure:Classifier}).
In addition, these networks respond rapidly: for most initial conditions, trajectories converge to the vicinity of the desired fixed points with less than 10 time units. Note that the associated topology map and the reactant--product incidence map are give in Extended Data Fig. \ref{EDF:Classifier}A).

The generated networks consistently favor certain reactions and exhibit specific patterns, as indicated by the reactant-product incidence map and representative circuits in \Cref{figure:Classifier}, demonstrating underlying design principles for this task.  
As in the previous tasks, the representative networks are chosen from across the clusters of high-performing candidates, where clusters are defined by topological similarity (see \Cref{figure:Classifier} bottom left).
These representative networks can be viewed as two independent branches---one for each species---combined via a sequestration reaction \ce{\species{X_1} + \species{X_2} -> $\emptyset$}. 
This reaction has a relatively high rate constant and acts as a competing module, driving the non-dominant species to approximately zero at steady state and thereby enabling the classification functionality.
Moreover, in each branch, an autoregulatory mechanism is provided to regulate the steady-state value of its associated species when it is dominant. 

After generation, we perform a post-design generalization analysis for the classifier circuits. Specifically, in addition to the 32 labeled initial conditions used in the design objective, we simulate each 40 top-ranked I/O CRN on an expanded set of 90 initial conditions, some of which closer to the target decision boundary (Fig.~\ref{figure:Classifier}A, bottom). We then recompute the classification loss on this expanded set and compare it to the objective loss, providing a direct check that the learned basin geometry extends beyond the discrete initial conditions used during the design process. This postprocessing step does not affect the design loop; it is used only to assess how well the generated fate-decision mechanisms interpolate to nearby initial states.

\begin{figure*}[ht!]
\includegraphics[scale=0.98]{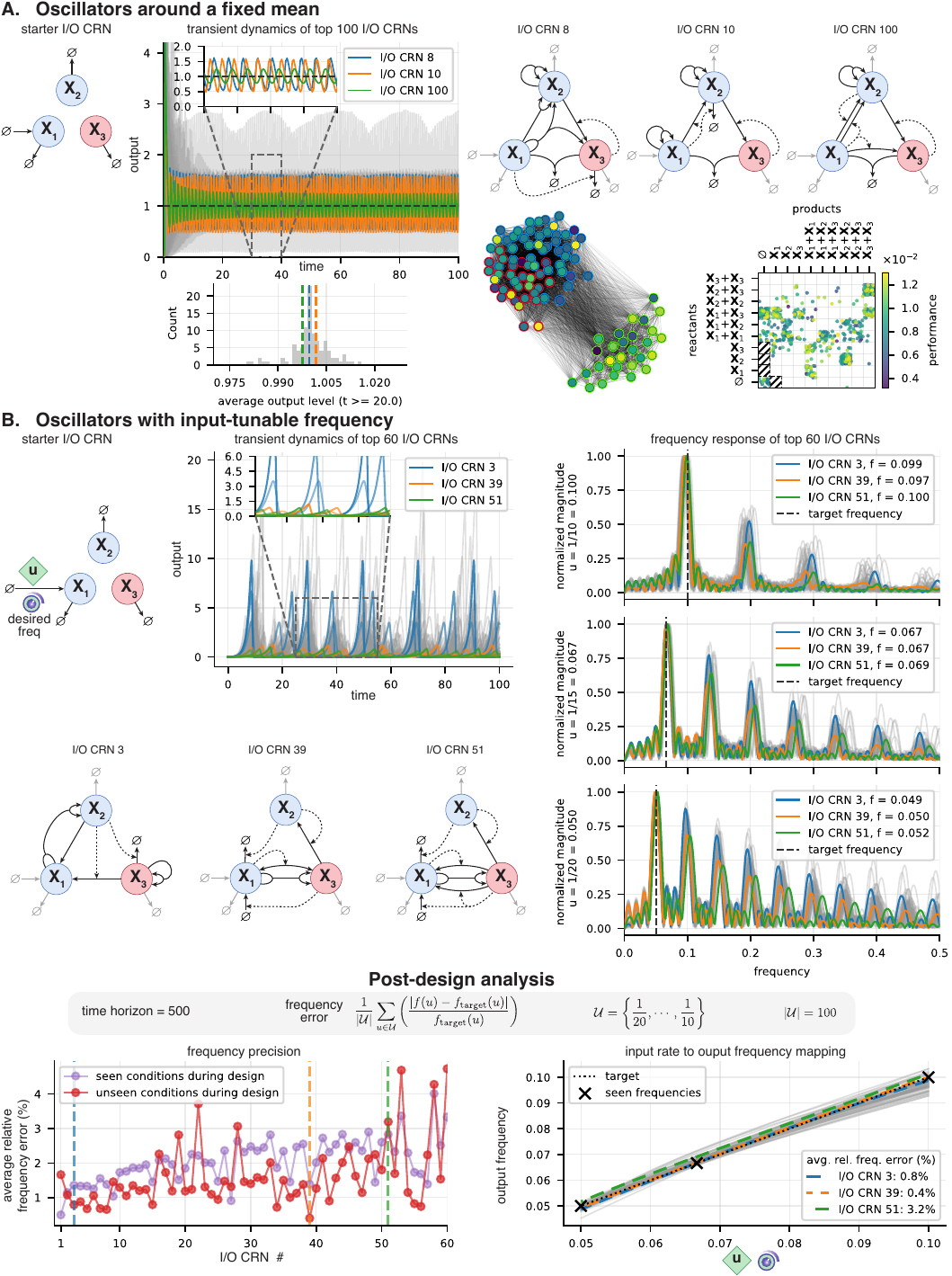}
\caption{\textbf{Molecular oscillators.} The figure caption is on the next page.}
\end{figure*}

\begin{figure*}[ht!]\ContinuedFloat
\caption{\textbf{Molecular oscillators.}
\textbf{(a) Oscillators around a fixed mean.} \crngen begins with a \startercrn (left) with three species, \species{X_1}, \species{X_2} (blue), and output \species{X_3} (red), and no external input. From this template, \crngen generates candidate I/O CRNs by appending up to 6 doubly bimolecular reactions, following mass-action kinetics, and optimizing a task objective that rewards sustained oscillations while constraining the temporal average output to a fixed mean equal to one (see loss function in Methods for more details). The center panel shows output trajectories for the top 100 topologically unique generated I/O CRNs (thin gray curves), with three representative solutions highlighted (colored; I/O CRNs 8, 10, and 100). The histogram (bottom left) summarizes the distribution of time-averaged output levels across the 100 solutions (computed over $t \ge 20$), illustrating concentration near the target mean. Example I/O CRN topologies for the highlighted networks are shown to the right. The graph visualization (bottom center) summarizes topological diversity across the solution set, and the reactant--product incidence map (bottom right) summarizes reaction usage across generated solutions.
\textbf{(b) Oscillators with input-tunable frequency.} A related \startercrn introduces an external input $u_1$ (green) that specifies a \emph{target oscillation frequency}. \crngen\ again appends up to 6 reactions and optimizes a frequency-tracking objective (see loss function in Methods for details), producing I/O CRNs whose oscillation frequency is tuned by $u_1$ while maintaining oscillatory behavior. The left-center plot shows time-domain output trajectories for the top 60 topologically unique solutions (gray), with three representative networks highlighted (I/O CRNs 1, 2, and 51). Frequency-domain behavior is summarized on the right: for three example input settings, the normalized magnitude spectra of the highlighted networks are shown, with dashed lines indicating the target frequencies; legend annotations report the achieved dominant frequencies $f$ for each I/O CRN. Example network topologies for the highlighted solutions are shown beneath the time-domain plot, and the topological diversity graph and reactant--product incidence map (bottom) summarize structural diversity and reaction usage across the generated oscillator solution set. 
{\bf Post-design testing beyond seen conditions.} To assess generalization, the top I/O CRNs from (B) are evaluated on a dense set of 100 input frequencies spanning the design range (schematic, top), using a longer simulation horizon ($t_{\mathrm{f}}=500$). The left panel compares the average relative frequency error on the input frequencies used during design (purple) versus additional frequencies not used during design (red) across the top 60 I/O CRNs. Using the longer horizon improves frequency estimation and yields higher precision, including on unseen input frequencies. The right panel plots output frequency versus input frequency for all top 60 I/O CRNs (gray) and the representative networks (colored), showing close agreement with the identity line and reporting the average relative error for each highlighted I/O CRN.}
\label{fig:Oscillators}
\noindent\rule{\linewidth}{0.4pt}
\end{figure*}

\subsection*{Logic circuits}
Logic circuits provide a compact and compositional language for programming biochemical systems: by mapping molecular inputs to discrete decisions, they enable modular information processing and serve as reusable building blocks for constructing more complex computations (e.g., multi-layer decision cascades, context-dependent responses, and higher-order CRN ``programs'' that orchestrate dynamics across many species) \cite{cherry2018scaling, moon2012genetic, seelig2006enzyme}.

With a \startercrn in which four external inputs $u_1,\cdots,u_4$ take Boolean values (low/high; $0/1$) and modulate the corresponding input species \species{X_1}$,\cdots,$\species{X_4} to regulate an output species \species{X_5}, \crngen\ discovers molecular circuits that implement the irreducible 4-input Boolean function $(u_1 \wedge u_2)\ \vee\ (u_2 \wedge u_3)\ \vee\ (u_3 \wedge u_4)$ (Fig. \ref{figure:Classifier}B). Allowing six appended doubly bimolecular mass-action reactions yields 29 top-performing I/O CRNs whose output trajectories cleanly separate into low and high digital levels. We highlight the transient dynamics of three representative solutions (I/O CRNs 1, 2, and 4; colored curves) and mark the low/high setpoints with horizontal dashed lines (Fig. \ref{figure:Classifier}B).

These solutions share two recurring design features. First, the output is stabilized by an \emph{autocatalytic degradation} motif, which suppresses overshoot and accelerates convergence. Second, the logic is computed in an \emph{analog} manner through direct or catalyzed conversion channels in which pairs of input species promote production of \species{X_5}, encoding the conjunctive clauses of the target function (Fig. \ref{figure:Classifier}B, example topologies). Tight binarization across all 16 input combinations is achieved by adding \emph{catalytic degradation} and \emph{interconversion} reactions among \species{X_1},$\cdots$,\species{X_4} that tune the effective high-state setpoint, so true cases saturate near unity rather than increasing with input magnitude (Fig. \ref{figure:Classifier}B), truth tables for I/O CRNs 1, 2, and 4). 
Note that the associated topology map and the reactant--product incidence map are give in Extended Data Fig. \ref{EDF:Classifier}B).

Restricting the search to at most five appended reactions yields 40 top-ranked candidates, and we show representative topologies (I/O CRNs 1, 6, and 26; Extended Data Fig. \ref{EDF:Classifier}C). With this reduced reaction budget, the setpoint-conditioning mechanisms above are typically absent, leading to a systematic overshoot above the nominal high level in the all-inputs-high condition (Extended Data Fig.~\ref{EDF:Classifier}C, time courses). This overshoot does not affect threshold-based classification, but it reveals a trade-off between circuit compactness and tight digital saturation that is alleviated by the sixth reaction in Fig.~\ref{figure:Classifier}B.

Finally, to demonstrate \crngen’s ability to operate on reaction libraries with multi-parameter kinetics, we apply it to a Hill-type production library and successfully generate diverse I/O CRNs, as shown in Extended Data Fig.~\ref{EDF:HF}.

\subsection*{\crngen\ provides biochemical oscillators with specific centers and tunable frequencies}

Biological rhythms are widespread in biology, precisely regulating life processes in response to oscillatory environmental cues.
Despite the discovery of natural biochemical oscillators \cite{vitaterna1994mutagenesis,bargiello1984molecular,geva2006oscillations,chen2015emergent,chandra2011glycolytic}, and the development of synthetic ones \cite{elowitz2000synthetic,tigges2010synthetic,novak2008design,kim2011synthetic}, designing circuits with specified or tunable oscillatory parameters (frequency, center, amplitude, etc.) remains prohibitively challenging, with only a few rational designs available \cite{stricker2008fast,qin2021frequency}.
We therefore tasked \crngen\ with constructing such biochemical oscillators from a \startercrn comprising three species and four reactions where all three species are diluted (top-left panels in \Cref{fig:Oscillators}A--B).

We begin with oscillators whose trajectories are required to oscillate around a prescribed center (Fig.~\ref{fig:Oscillators}A). In this setting the \startercrn has no inputs, and the objective targets the mean (center) of the output oscillation. \crngen\ discovers more than 100 topologically distinct high-performing networks, with oscillator centers tightly clustered around the target value (Fig.~\ref{fig:Oscillators}A, histogram). Three representative solutions are highlighted with colored trajectories alongside their corresponding topologies, while the topological diversity graph summarizes the breadth of distinct solutions found by \crngen. Interestingly, the sequestration reaction \ce{\species{X_1} + \species{X_3} -> $\emptyset$} emerges repeatedly in the generated CRNs (see the table in Fig.~\ref{fig:Oscillators}A), suggesting that sequestration constitutes a structurally favored mechanism for oscillatory regulation. Notably, related sequestration dynamics play a central role in the mmammalian circadian clock \cite{kim2012mechanism, jeong2022combined}. We are grateful to Prof. Jae Kyoung Kim for drawing our attention to this connection.

We next consider oscillators with \emph{input-tunable} frequency (Fig.~\ref{fig:Oscillators}B). Here the \startercrn includes an input $u_1$ that modulates the production rate of \species{X_1}, providing a tuning knob for the oscillation frequency of the output species \species{X_3}. We evaluate three target periods ($T=10,15,20$) and so the input $u_1$ takes 3 values $u_1 \in \{1/20, 1/15, 1/10\}$, and find that \crngen\ again produces more than 60 high-performance networks (Fig.~\ref{fig:Oscillators}B). Fourier spectra (Fig.~\ref{fig:Oscillators}B, right) confirm that, for each input level, the fundamental frequency aligns with the prescribed target period. As in the previous task, the diversity graph and reactant--product incidence map reveal substantial topological variation and recurring reaction usage among the generated solutions (see Extended Data Fig.~\ref{EDF:Oscillators}), highlighting the framework’s ability to generate multiple distinct implementations of the same dynamical specification.

After generation, we perform a post-design generalization test for the frequency-tunable oscillators. Specifically, although the design objective in Fig.~\ref{fig:Oscillators}B evaluates only a sparse set of target frequencies, we take the top I/O CRNs and simulate them on a dense sweep of 100 input frequencies spanning the same range. To improve frequency estimation, we use a longer simulation horizon ($t_{\mathrm{f}}=500$), which yields higher-precision frequency measurements and correspondingly lower relative error, including for input frequencies not encountered during design. The resulting input–output frequency curves closely track the identity relation, indicating that many generated oscillators interpolate smoothly across the design range rather than merely matching the discrete targets.

\section*{Discussion}

We presented \crngen, a generative AI framework for the automated design of chemical reaction networks from high-level dynamical specifications. \crngen\ exploits the asymmetry between the ease of \emph{evaluating} a proposed CRN by simulation and the difficulty of \emph{discovering} one that satisfies a target behavior. By placing an AI agent in the loop, \crngen\ iteratively proposes reactions (including kinetic parameters and, when relevant, input influences), evaluates completed networks with a task-defined loss, and refines its proposal distribution to generate diverse, high-performing solutions.

Across a broad set of benchmarks—including dose–response shaping (Hill-type, Michaelis–Menten, ultrasensitive, and non-monotonic profiles), logic circuits implementing complex Boolean functions, classification by initial conditions, oscillators with constrained mean or input-tunable frequency, robust perfect adaptation under deterministic and stochastic dynamics (including simultaneous noise reduction), and habituation/sensitization—\crngen\ consistently produced families of topologically distinct networks that match the desired behaviors. Beyond performance, the resulting solution sets enabled mechanistic insight: recurring reaction patterns and cluster-level structure revealed alternative motifs for achieving the same specification.

Despite all these successes, several directions remain to be explored in future work to further improve the method and extend its applicability to more advanced biological studies.
From a methodological perspective, the computational efficiency of our generative approach could potentially be improved by adopting more advanced neural network architectures and generative paradigms, such as transformer-based models \cite{vaswani2017attention}, GFlowNets \cite{zhang2023let}, and diffusion models \cite{DBLP:journals/corr/abs-2006-11239}.
In addition, incorporating more advanced training strategies—such as Proximal Policy Optimization (PPO) \cite{schulman2017proximal}—may further accelerate learning and yield more robust performance.
Moreover, an adaptive hyperparameter-tuning scheme would help streamline practical use and reduce the burden of manual calibration for end users.

Our current implementation is designed to run on a single computer node (workstation or cluster job); however, because candidate CRNs and evaluation conditions are largely independent, the workflow is inherently parallel. Extending \crngen\ to distributed, multi-node execution is therefore a natural next step, but it would require careful systems engineering (e.g., asynchronous scheduling of simulations, efficient aggregation of rewards/gradients) to correctly handle the distributed training of our agent.

On the other hand, \crngen\ naturally extends to richer design settings. One immediate direction is to treat \emph{input assignment} itself as part of the search, allowing the agent to decide not only which reactions to add but also which external signals modulate each propensity. 
Although we focused here on reaction libraries with mass-action kinetics and hill kinetics (see Extended Data Fig.~\ref{EDF:HF}), \crngen\ is not limited to these settings. The same framework can operate with arbitrary, user-defined reaction libraries in which each reaction carries multiple kinetic parameters and follows more complex propensity forms.
The framework can also leverage more sophisticated simulation environments that incorporate additional biological context—such as growth, dilution, resource limitations, burden, spatial effects, or host–circuit interactions—so that designs are optimized under more realistic operating conditions. Beyond the benchmarks explored here, the same loop could be applied to a wider range of objectives and larger networks, including multi-module behaviors and multi-objective specifications. Finally, a complementary avenue is to decouple structure and parameters more explicitly: the agent could focus primarily on proposing reaction \emph{topologies}, while an embedded optimizer (or inner-loop parameter-fitting routine) searches for the best kinetic parameters for each proposed topology, potentially improving sample efficiency and sharpening the topology-level search.

\crngen\ is intended as a general design engine rather than a solver for a single task class. Users can specify a \startercrn, a reaction library, kinetic semantics, and an evaluation objective, allowing the same framework to be applied across modeling assumptions and application domains. Looking forward, \crngen\ opens a path toward more programmable molecular engineering, where circuit discovery is guided directly by behavioral specifications and where interpretable reaction-level motifs can be rapidly surfaced, compared, and translated into experimental implementations.

\begin{acknowledgements}
\noindent
This work is supported by the Swiss State Secretariat for Education, Research and Innovation (SERI).
Z.F. is supported by the National Natural Science Foundation of China (Grant No. 12501699) and the Open Foundation of the State Key Laboratory of Mathematical Sciences (Grant No. SLMS-2025-KFKT-TD-01).
\end{acknowledgements}

\section*{References}
\bibliography{zHenriquesLab-Mendeley}

@article{malekpour2023wplogicnet,
  title={wpLogicNet: logic gate and structure inference in gene regulatory networks},
  author={Malekpour, Seyed Amir and Shahdoust, Maryam and Aghdam, Rosa and Sadeghi, Mehdi},
  journal={Bioinformatics},
  volume={39},
  number={2},
  pages={btad072},
  year={2023},
  publisher={Oxford University Press}
}

@article{kim2011synthetic,
  title={Synthetic in vitro transcriptional oscillators},
  author={Kim, Jongmin and Winfree, Erik},
  journal={Molecular systems biology},
  volume={7},
  number={1},
  pages={465},
  year={2011},
  publisher={John Wiley \& Sons, Ltd Chichester, UK}
}

@article{seelig2006enzyme,
  title={Enzyme-free nucleic acid logic circuits},
  author={Seelig, Georg and Soloveichik, David and Zhang, David Yu and Winfree, Erik},
  journal={science},
  volume={314},
  number={5805},
  pages={1585--1588},
  year={2006},
  publisher={American Association for the Advancement of Science}
}

@article{tigges2010synthetic,
  title={A synthetic low-frequency mammalian oscillator},
  author={Tigges, Marcel and D{\'e}nervaud, Nicolas and Greber, David and Stelling, Joerg and Fussenegger, Martin},
  journal={Nucleic acids research},
  volume={38},
  number={8},
  pages={2702--2711},
  year={2010},
  publisher={Oxford University Press}
}

@article{novak2008design,
  title={Design principles of biochemical oscillators},
  author={Nov{\'a}k, B{\'e}la and Tyson, John J},
  journal={Nature reviews Molecular cell biology},
  volume={9},
  number={12},
  pages={981--991},
  year={2008},
  publisher={Nature Publishing Group UK London}
}

@article{chandra2011glycolytic,
  title={Glycolytic oscillations and limits on robust efficiency},
  author={Chandra, Fiona A and Buzi, Gentian and Doyle, John C},
  journal={science},
  volume={333},
  number={6039},
  pages={187--192},
  year={2011},
  publisher={American Association for the Advancement of Science}
}

@article{qin2021frequency,
  title={A frequency-amplitude coordinator and its optimal energy consumption for biological oscillators},
  author={Qin, Bo-Wei and Zhao, Lei and Lin, Wei},
  journal={Nature Communications},
  volume={12},
  number={1},
  pages={5894},
  year={2021},
  publisher={Nature Publishing Group UK London}
}

@article{chen2015emergent,
  title={Emergent genetic oscillations in a synthetic microbial consortium},
  author={Chen, Ye and Kim, Jae Kyoung and Hirning, Andrew J and Josi{\'c}, Kre{\v{s}}imir and Bennett, Matthew R},
  journal={Science},
  volume={349},
  number={6251},
  pages={986--989},
  year={2015},
  publisher={American Association for the Advancement of Science}
}

@article{stricker2008fast,
  title={A fast, robust and tunable synthetic gene oscillator},
  author={Stricker, Jesse and Cookson, Scott and Bennett, Matthew R and Mather, William H and Tsimring, Lev S and Hasty, Jeff},
  journal={Nature},
  volume={456},
  number={7221},
  pages={516--519},
  year={2008},
  publisher={Nature Publishing Group UK London}
}

@article{geva2006oscillations,
  title={Oscillations and variability in the p53 system},
  author={Geva-Zatorsky, Naama and Rosenfeld, Nitzan and Itzkovitz, Shalev and Milo, Ron and Sigal, Alex and Dekel, Erez and Yarnitzky, Talia and Liron, Yuvalal and Polak, Paz and Lahav, Galit and others},
  journal={Molecular systems biology},
  volume={2},
  number={1},
  pages={2006--0033},
  year={2006},
  publisher={John Wiley \& Sons, Ltd Chichester, UK}
}

@article{bargiello1984molecular,
  title={Molecular genetics of a biological clock in Drosophila},
  author={Bargiello, Thaddeus A and Young, Michael W},
  journal={Proceedings of the National Academy of Sciences},
  volume={81},
  number={7},
  pages={2142--2146},
  year={1984}
}

@article{vitaterna1994mutagenesis,
  title={Mutagenesis and mapping of a mouse gene, Clock, essential for circadian behavior},
  author={Vitaterna, Martha Hotz and King, David P and Chang, Anne-Marie and Kornhauser, Jon M and Lowrey, Phillip L and McDonald, J David and Dove, William F and Pinto, Lawrence H and Turek, Fred W and Takahashi, Joseph S},
  journal={Science},
  volume={264},
  number={5159},
  pages={719--725},
  year={1994},
  publisher={American Association for the Advancement of Science}
}

@article{rai2024using,
  title={Using machine learning to enhance and accelerate synthetic biology},
  author={Rai, Kshitij and Wang, Yiduo and O'Connell, Ronan W and Patel, Ankit B and Bashor, Caleb J},
  journal={Current Opinion in Biomedical Engineering},
  volume={31},
  pages={100553},
  year={2024},
  publisher={Elsevier}
}

@article{merzbacher2023bayesian,
  title={Bayesian optimization for design of multiscale biological circuits},
  author={Merzbacher, Charlotte and Mac Aodha, Oisin and Oyarz{\'u}n, Diego A},
  journal={ACS synthetic biology},
  volume={12},
  number={7},
  pages={2073--2082},
  year={2023},
  publisher={ACS Publications}
}

@article{palacios2025machine,
  title={Machine learning for synthetic gene circuit engineering},
  author={Palacios, Sebastian and Collins, James J and Del Vecchio, Domitilla},
  journal={Current Opinion in Biotechnology},
  volume={92},
  pages={103263},
  year={2025},
  publisher={Elsevier}
}

@article{rybinski2020topofilter,
  title={TopoFilter: a MATLAB package for mechanistic model identification in systems biology},
  author={Rybi{\'n}ski, Miko{\l}aj and M{\"o}ller, Simon and Sunn{\aa}ker, Mikael and Lormeau, Claude and Stelling, J{\"o}rg},
  journal={BMC bioinformatics},
  volume={21},
  number={1},
  pages={34},
  year={2020},
  publisher={Springer}
}

@article{roybal2016precision,
  title={Precision tumor recognition by {T} cells with combinatorial antigen-sensing circuits},
  author={Roybal, Kole T and Rupp, Levi J and Morsut, Leonardo and Walker, Whitney J and McNally, Krista A and Park, Jason S and Lim, Wendell A},
  journal={Cell},
  volume={164},
  number={4},
  pages={770--779},
  year={2016},
  publisher={Elsevier}
}

@article{porter2011chimeric,
  title={Chimeric antigen receptor--modified T cells in chronic lymphoid leukemia},
  author={Porter, David L and Levine, Bruce L and Kalos, Michael and Bagg, Adam and June, Carl H},
  journal={New England Journal of Medicine},
  volume={365},
  number={8},
  pages={725--733},
  year={2011},
  publisher={Mass Medical Soc}
}

@article{kosuri2014large,
  title={Large-scale de novo {DNA} synthesis: technologies and applications},
  author={Kosuri, Sriram and Church, George M},
  journal={Nature methods},
  volume={11},
  number={5},
  pages={499--507},
  year={2014},
  publisher={Nature Publishing Group US New York}
}

@article{cong2013multiplex,
  title={Multiplex genome engineering using {CRISPR/Cas} systems},
  author={Cong, Le and Ran, F Ann and Cox, David and Lin, Shuailiang and Barretto, Robert and Habib, Naomi and Hsu, Patrick D and Wu, Xuebing and Jiang, Wenyan and Marraffini, Luciano A and others},
  journal={Science},
  volume={339},
  number={6121},
  pages={819--823},
  year={2013},
  publisher={American Association for the Advancement of Science}
}

@article{jinek2012programmable,
  title={A programmable dual-RNA--guided {DNA} endonuclease in adaptive bacterial immunity},
  author={Jinek, Martin and Chylinski, Krzysztof and Fonfara, Ines and Hauer, Michael and Doudna, Jennifer A and Charpentier, Emmanuelle},
  journal={science},
  volume={337},
  number={6096},
  pages={816--821},
  year={2012},
  publisher={American Association for the Advancement of Science}
}

@inproceedings{rossi2024synthevo,
  title={SynthEvo: A Gradient-Guided Evolutionary Approach for Synthetic Circuit Design},
  author={Rossi, Nicol{\'o} and Gupta, Ankit and Khammash, Mustafa},
  booktitle={2024 IEEE 63rd Conference on Decision and Control (CDC)},
  pages={6298--6304},
  year={2024},
  organization={IEEE}
}

@inproceedings{moorman2019dynamical,
  title={A dynamical biomolecular neural network},
  author={Moorman, Andrew and Samaniego, Christian Cuba and Maley, Carlo and Weiss, Ron},
  booktitle={2019 IEEE 58th conference on decision and control (CDC)},
  pages={1797--1802},
  year={2019},
  organization={IEEE}
}

@article{oishi2011biomolecular,
  title={Biomolecular implementation of linear I/O systems},
  author={Oishi, Kevin and Klavins, Eric},
  journal={IET systems biology},
  volume={5},
  number={4},
  pages={252--260},
  year={2011},
  publisher={IET}
}

@article{okumura2022nonlinear,
  title={Nonlinear decision-making with enzymatic neural networks},
  author={Okumura, Shu and Gines, Guillaume and Lobato-Dauzier, Nicolas and Baccouche, Alexandre and Deteix, Robin and Fujii, Teruo and Rondelez, Yannick and Genot, Anthony J},
  journal={Nature},
  volume={610},
  number={7932},
  pages={496--501},
  year={2022},
  publisher={Nature Publishing Group UK London}
}

@article{cherry2018scaling,
  title={Scaling up molecular pattern recognition with {DNA}-based winner-take-all neural networks},
  author={Cherry, Kevin M and Qian, Lulu},
  journal={Nature},
  volume={559},
  number={7714},
  pages={370--376},
  year={2018},
  publisher={Nature Publishing Group UK London}
}

@article{vasic2022programming,
  title={Programming and training rate-independent chemical reaction networks},
  author={Vasi{\'c}, Marko and Chalk, Cameron and Luchsinger, Austin and Khurshid, Sarfraz and Soloveichik, David},
  journal={Proceedings of the National Academy of Sciences},
  volume={119},
  number={24},
  pages={e2111552119},
  year={2022},
  publisher={National Academy of Sciences}
}

@article{zechner2016molecular,
  title={Molecular circuits for dynamic noise filtering},
  author={Zechner, Christoph and Seelig, Georg and Rullan, Marc and Khammash, Mustafa},
  journal={Proceedings of the National Academy of Sciences},
  volume={113},
  number={17},
  pages={4729--4734},
  year={2016},
  publisher={National Academy of Sciences}
}

@article{whitby2021pid,
  title={PID control of biochemical reaction networks},
  author={Whitby, Max and Cardelli, Luca and Kwiatkowska, Marta and Laurenti, Luca and Tribastone, Mirco and Tschaikowski, Max},
  journal={IEEE Transactions on Automatic Control},
  volume={67},
  number={2},
  pages={1023--1030},
  year={2021},
  publisher={IEEE}
}

@article{paulino2019pid,
  title={PID and state feedback controllers using {DNA} strand displacement reactions},
  author={Paulino, Nuno MG and Foo, Mathias and Kim, Jongmin and Bates, Declan G},
  journal={IEEE Control Systems Letters},
  volume={3},
  number={4},
  pages={805--810},
  year={2019},
  publisher={IEEE}
}

@article{alexis2022design,
  title={On the design of a PID bio-controller with set point weighting and filtered derivative action},
  author={Alexis, Emmanouil and Cardelli, Luca and Papachristodoulou, Antonis},
  journal={IEEE Control Systems Letters},
  volume={6},
  pages={3134--3139},
  year={2022},
  publisher={IEEE}
}

@article{chevalier2019design,
  title={Design and analysis of a proportional-integral-derivative controller with biological molecules},
  author={Chevalier, Michael and G{\'o}mez-Schiavon, Mariana and Ng, Andrew H and El-Samad, Hana},
  journal={Cell systems},
  volume={9},
  number={4},
  pages={338--353},
  year={2019},
  publisher={Elsevier}
}

@article{jones2022robust,
  title={Robust and tunable signal processing in mammalian cells via engineered covalent modification cycles},
  author={Jones, Ross D and Qian, Yili and Ilia, Katherine and Wang, Benjamin and Laub, Michael T and Del Vecchio, Domitilla and Weiss, Ron},
  journal={Nature communications},
  volume={13},
  number={1},
  pages={1720},
  year={2022},
  publisher={Nature Publishing Group UK London}
}

@article{anastassov2023cybergenetic,
  title={A cybergenetic framework for engineering intein-mediated integral feedback control systems},
  author={Anastassov, Stanislav and Filo, Maurice and Chang, Ching-Hsiang and Khammash, Mustafa},
  journal={Nature Communications},
  volume={14},
  number={1},
  pages={1337},
  year={2023},
  publisher={Nature Publishing Group UK London}
}

@article{briat2016antithetic,
  title={Antithetic integral feedback ensures robust perfect adaptation in noisy biomolecular networks},
  author={Briat, Corentin and Gupta, Ankit and Khammash, Mustafa},
  journal={Cell systems},
  volume={2},
  number={1},
  pages={15--26},
  year={2016},
  publisher={Elsevier}
}

@article{kell2023noise,
  title={Noise properties of adaptation-conferring biochemical control modules},
  author={Kell, Brayden and Ripsman, Ryan and Hilfinger, Andreas},
  journal={Proceedings of the National Academy of Sciences},
  volume={120},
  number={38},
  pages={e2302016120},
  year={2023},
  publisher={National Academy of Sciences}
}

@article{khalil2010synthetic,
  title={Synthetic biology: applications come of age},
  author={Khalil, Ahmad S and Collins, James J},
  journal={Nature Reviews Genetics},
  volume={11},
  number={5},
  pages={367--379},
  year={2010},
  publisher={Nature Publishing Group UK London}
}

@article{voigt2020synthetic,
  title={Synthetic biology 2020--2030: six commercially-available products that are changing our world},
  author={Voigt, Christopher A},
  journal={Nature Communications},
  volume={11},
  number={1},
  pages={6379},
  year={2020},
  publisher={Nature Publishing Group UK London}
}

@article{fan2025automatic,
  title={Automatic Implementation of Neural Networks through Reaction Networks—Part I: Circuit Design and Convergence Analysis},
  author={Fan, Yuzhen and Zhang, Xiaoyu and Gao, Chuanhou and Dochain, Denis},
  journal={IEEE Transactions on Automatic Control},
  year={2025},
  publisher={IEEE}
}

@article{chen2024synthetic,
  title={A synthetic protein-level neural network in mammalian cells},
  author={Chen, Zibo and Linton, James M and Xia, Shiyu and Fan, Xinwen and Yu, Dingchen and Wang, Jinglin and Zhu, Ronghui and Elowitz, Michael B},
  journal={Science},
  volume={386},
  number={6727},
  pages={1243--1250},
  year={2024},
  publisher={American Association for the Advancement of Science}
}

@article{filo2022hierarchy,
  title={A hierarchy of biomolecular proportional-integral-derivative feedback controllers for robust perfect adaptation and dynamic performance},
  author={Filo, Maurice and Kumar, Sant and Khammash, Mustafa},
  journal={Nature communications},
  volume={13},
  number={1},
  pages={2119},
  year={2022},
  publisher={Nature Publishing Group UK London}
}

@article{frei2022genetic,
  title={A genetic mammalian proportional--integral feedback control circuit for robust and precise gene regulation},
  author={Frei, Timothy and Chang, Ching-Hsiang and Filo, Maurice and Arampatzis, Asterios and Khammash, Mustafa},
  journal={Proceedings of the National Academy of Sciences},
  volume={119},
  number={00},
  pages={e2122132119},
  year={2022},
  publisher={National Academy of Sciences}
}

@article{huang2018quasi,
  title={A quasi-integral controller for adaptation of genetic modules to variable ribosome demand},
  author={Huang, Hsin-Ho and Qian, Yili and Del Vecchio, Domitilla},
  journal={Nature communications},
  volume={9},
  number={1},
  pages={5415},
  year={2018},
  publisher={Nature Publishing Group UK London}
}

@article{becskei2000engineering,
  title={Engineering stability in gene networks by autoregulation},
  author={Becskei, Attila and Serrano, Luis},
  journal={Nature},
  volume={405},
  number={6786},
  pages={590--593},
  year={2000},
  publisher={Nature Publishing Group UK London}
}

@article{aoki2019universal,
  title={A universal biomolecular integral feedback controller for robust perfect adaptation},
  author={Aoki, Stephanie K and Lillacci, Gabriele and Gupta, Ankit and Baumschlager, Armin and Schweingruber, David and Khammash, Mustafa},
  journal={Nature},
  volume={570},
  number={7762},
  pages={533--537},
  year={2019},
  publisher={Nature Publishing Group UK London}
}

@article{gardner2000construction,
  title={Construction of a genetic toggle switch in Escherichia coli},
  author={Gardner, Timothy S and Cantor, Charles R and Collins, James J},
  journal={Nature},
  volume={403},
  number={6767},
  pages={339--342},
  year={2000},
  publisher={Nature Publishing Group UK London}
}

@article{elowitz2000synthetic,
  title={A synthetic oscillatory network of transcriptional regulators},
  author={Elowitz, Michael B and Leibler, Stanislas},
  journal={Nature},
  volume={403},
  number={6767},
  pages={335--338},
  year={2000},
  publisher={Nature Publishing Group UK London}
}

@article{gallup2025generative,
  title={Generative design of synthetic gene circuits for functional and evolutionary properties},
  author={Gallup, Olivia and Steel, Harrison},
  journal={bioRxiv},
  pages={2025--09},
  year={2025},
  publisher={Cold Spring Harbor Laboratory}
}

@article{ma2009defining,
  title={Defining network topologies that can achieve biochemical adaptation},
  author={Ma, Wenzhe and Trusina, Ala and El-Samad, Hana and Lim, Wendell A and Tang, Chao},
  journal={Cell},
  volume={138},
  number={4},
  pages={760--773},
  year={2009},
  publisher={Elsevier}
}

@article{hiscock2019adapting,
  title={Adapting machine-learning algorithms to design gene circuits},
  author={Hiscock, Tom W},
  journal={BMC bioinformatics},
  volume={20},
  number={1},
  pages={214},
  year={2019},
  publisher={Springer}
}

@article{kobiela2024risk,
  title={Risk-averse optimization of genetic circuits under uncertainty},
  author={Kobiela, Michal and Oyarz{\'u}n, Diego A and Gutmann, Michael U},
  journal={bioRxiv},
  pages={2024--11},
  year={2024},
  publisher={Cold Spring Harbor Laboratory}
}

@article{bhamidipati2025designing,
  title={Designing biochemical circuits with tree search},
  author={Bhamidipati, Pranav S and Thomson, Matthew},
  journal={bioRxiv},
  pages={2025--01},
  year={2025},
  publisher={Cold Spring Harbor Laboratory}
}

@inproceedings{samaniego2024neural,
  title={Neural networks built from enzymatic reactions can operate as linear and nonlinear classifiers},
  author={Samaniego, Christian Cuba and Wallace, Emily and Blanchini, Franco and Franco, Elisa and Giordano, Giulia},
  booktitle={2024 IEEE 63rd Conference on Decision and Control (CDC)},
  pages={6292--6297},
  year={2024},
  organization={IEEE}
}

@article{anderson2021reaction,
  title={On reaction network implementations of neural networks},
  author={Anderson, David F and Joshi, Badal and Deshpande, Abhishek},
  journal={Journal of the Royal Society Interface},
  volume={18},
  number={177},
  pages={20210031},
  year={2021},
  publisher={The Royal Society}
}

@article{cherry2025supervised,
  title={Supervised learning in {DNA} neural networks},
  author={Cherry, Kevin M and Qian, Lulu},
  journal={Nature},
  pages={1--9},
  year={2025},
  publisher={Nature Publishing Group UK London}
}

@article{moon2012genetic,
  title={Genetic programs constructed from layered logic gates in single cells},
  author={Moon, Tae Seok and Lou, Chunbo and Tamsir, Alvin and Stanton, Brynne C and Voigt, Christopher A},
  journal={Nature},
  volume={491},
  number={7423},
  pages={249--253},
  year={2012},
  publisher={Nature Publishing Group UK London}
}

@article{zhang2023let,
  title={Let the flows tell: Solving graph combinatorial problems with gflownets},
  author={Zhang, Dinghuai and Dai, Hanjun and Malkin, Nikolay and Courville, Aaron C and Bengio, Yoshua and Pan, Ling},
  journal={Advances in neural information processing systems},
  volume={36},
  pages={11952--11969},
  year={2023}
}

@article{schulman2017proximal,
  title={Proximal policy optimization algorithms},
  author={Schulman, John and Wolski, Filip and Dhariwal, Prafulla and Radford, Alec and Klimov, Oleg},
  journal={arXiv preprint arXiv:1707.06347},
  year={2017}
}

@article{williams1992simple,
  title={Simple statistical gradient-following algorithms for connectionist reinforcement learning},
  author={Williams, Ronald J},
  journal={Machine learning},
  volume={8},
  number={3},
  pages={229--256},
  year={1992},
  publisher={Springer}
}

@article{chow2014algorithms,
  title={Algorithms for CVaR optimization in MDPs},
  author={Chow, Yinlam and Ghavamzadeh, Mohammad},
  journal={Advances in neural information processing systems},
  volume={27},
  year={2014}
}

@inproceedings{geist2019theory,
  title={A theory of regularized markov decision processes},
  author={Geist, Matthieu and Scherrer, Bruno and Pietquin, Olivier},
  booktitle={International conference on machine learning},
  pages={2160--2169},
  year={2019},
  organization={PMLR}
}

@inproceedings{oh2018self,
  title={Self-imitation learning},
  author={Oh, Junhyuk and Guo, Yijie and Singh, Satinder and Lee, Honglak},
  booktitle={International conference on machine learning},
  pages={3878--3887},
  year={2018},
  organization={PMLR}
}

@article{delalleau2019discrete,
  title={Discrete and continuous action representation for practical RL in video games},
  author={Delalleau, Olivier and Peter, Maxim and Alonso, Eloi and Logut, Adrien},
  journal={arXiv preprint arXiv:1912.11077},
  year={2019}
}

@article{filo2023biomolecular,
  title={Biomolecular feedback controllers: from theory to applications},
  author={Filo, Maurice and Chang, Ching-Hsiang and Khammash, Mustafa},
  journal={Current Opinion in Biotechnology},
  volume={79},
  pages={102882},
  year={2023},
  publisher={Elsevier}
}

@article{kotas2015homeostasis,
  title={Homeostasis, inflammation, and disease susceptibility},
  author={Kotas, Maya E and Medzhitov, Ruslan},
  journal={Cell},
  volume={160},
  number={5},
  pages={816--827},
  year={2015},
  publisher={Elsevier}
}

@article{khammash2021perfect,
  title={Perfect adaptation in biology},
  author={Khammash, Mustafa H},
  journal={Cell Systems},
  volume={12},
  number={6},
  pages={509--521},
  year={2021},
  publisher={Elsevier}
}

@article{frei2021adaptive,
  title={Adaptive circuits in synthetic biology},
  author={Frei, Timothy and Khammash, Mustafa},
  journal={Current Opinion in Systems Biology},
  volume={28},
  pages={100399},
  year={2021},
  publisher={Elsevier}
}

@inproceedings{xiao2018robust,
  title={Robust perfect adaptation in biomolecular reaction networks},
  author={Xiao, Fangzhou and Doyle, John C},
  booktitle={2018 IEEE conference on decision and control (CDC)},
  pages={4345--4352},
  year={2018},
  organization={IEEE}
}

@article{gupta2022universal,
  title={Universal structural requirements for maximal robust perfect adaptation in biomolecular networks},
  author={Gupta, Ankit and Khammash, Mustafa},
  journal={Proceedings of the National Academy of Sciences},
  volume={119},
  number={43},
  pages={e2207802119},
  year={2022},
  publisher={National Acad Sciences}
}

@article{gupta2023internal,
  title={The internal model principle for biomolecular control theory},
  author={Gupta, Ankit and Khammash, Mustafa},
  journal={IEEE Open Journal of Control Systems},
  volume={2},
  pages={63--69},
  year={2023},
  publisher={IEEE}
}

@article{hirono2025rethinking,
  title={Rethinking robust adaptation: Characterization of structural mechanisms for biochemical network robustness through topological invariants},
  author={Hirono, Yuji and Gupta, Ankit and Khammash, Mustafa},
  journal={PRX Life},
  volume={3},
  number={1},
  pages={013017},
  year={2025},
  publisher={APS}
}

@article{araujo2018topological,
  title={The topological requirements for robust perfect adaptation in networks of any size},
  author={Araujo, Robyn P and Liotta, Lance A},
  journal={Nature communications},
  volume={9},
  number={1},
  pages={1757},
  year={2018},
  publisher={Nature Publishing Group UK London}
}

@article{briat2016design,
  title={Design of a synthetic integral feedback circuit: dynamic analysis and {DNA} implementation},
  author={Briat, Corentin and Zechner, Christoph and Khammash, Mustafa},
  journal={ACS synthetic biology},
  volume={5},
  number={10},
  pages={1108--1116},
  year={2016},
  publisher={ACS Publications}
}

@article{ni2009control,
  title={The control of the controller: molecular mechanisms for robust perfect adaptation and temperature compensation},
  author={Ni, Xiao Yu and Drengstig, Tormod and Ruoff, Peter},
  journal={Biophysical journal},
  volume={97},
  number={5},
  pages={1244--1253},
  year={2009},
  publisher={Elsevier}
}

@article{drengstig2012robust,
  title={Robust adaptation and homeostasis by autocatalysis},
  author={Drengstig, T and Ni, XY and Thorsen, K and Jolma, IW and Ruoff, P},
  journal={The Journal of Physical Chemistry B},
  volume={116},
  number={18},
  pages={5355--5363},
  year={2012},
  publisher={ACS Publications}
}

@article{mallozzi2024crisprator,
  title={The {CRISPRaTOR}: a biomolecular circuit for Automatic Gene Regulation in Mammalian Cells with {CRISPR} technology.},
  author={Mallozzi, Alessio and Fusco, Virginia and Ragazzini, Francesco and di Bernardo, Diego},
  journal={bioRxiv},
  year={2024},
  publisher={Cold Spring Harbor Laboratory}
}

@article{gillespie1977exact,
  title={Exact stochastic simulation of coupled chemical reactions},
  author={Gillespie, Daniel T},
  journal={The journal of physical chemistry},
  volume={81},
  number={25},
  pages={2340--2361},
  year={1977},
  publisher={ACS Publications}
}

@article{filo2023hidden,
  title={A hidden proportional feedback mechanism underlies enhanced dynamic performance and noise rejection in sensor-based antithetic integral control},
  author={Filo, Maurice and Hou, Mucun and Khammash, M},
  journal={bioRxiv},
  pages={2023--04},
  year={2023},
  publisher={Cold Spring Harbor Laboratory}
}

@article{lim2025toward,
  title={Toward single-cell control: noise-robust perfect adaptation in biomolecular systems},
  author={Lim, Dongju and Moon, Seokhwan and Song, Yun Min and Kim, Minjun and Kim, Jinyeong and Kim, Kangsan and Cho, Byung-Kwan and Kim, Jinsu and Kim, Jae Kyoung},
  journal={Nature Communications},
  year={2025},
  publisher={Nature Publishing Group UK London}
}

@article{briat2018antithetic,
  title={Antithetic proportional-integral feedback for reduced variance and improved control performance of stochastic reaction networks},
  author={Briat, Corentin and Gupta, Ankit and Khammash, Mustafa},
  journal={Journal of The Royal Society Interface},
  volume={15},
  number={143},
  pages={20180079},
  year={2018},
  publisher={The Royal Society}
}

@article{hoose2023dna,
  title={{DNA} synthesis technologies to close the gene writing gap},
  author={Hoose, Alex and Vellacott, Richard and Storch, Marko and Freemont, Paul S and Ryadnov, Maxim G},
  journal={Nature Reviews Chemistry},
  volume={7},
  number={3},
  pages={144--161},
  year={2023},
  publisher={Nature Publishing Group UK London}
}

@article{chehelgerdi2024comprehensive,
  title={Comprehensive review of {CRISPR}-based gene editing: mechanisms, challenges, and applications in cancer therapy},
  author={Chehelgerdi, Mohammad and Chehelgerdi, Matin and Khorramian-Ghahfarokhi, Milad and Shafieizadeh, Marjan and Mahmoudi, Esmaeil and Eskandari, Fatemeh and Rashidi, Mohsen and Arshi, Asghar and Mokhtari-Farsani, Abbas},
  journal={Molecular cancer},
  volume={23},
  number={1},
  pages={9},
  year={2024},
  publisher={Springer}
}

@article{rodrigo2007genetdes,
  title={Genetdes: automatic design of transcriptional networks},
  author={Rodrigo, Guillermo and Carrera, Javier and Jaramillo, Alfonso},
  journal={Bioinformatics},
  volume={23},
  number={14},
  pages={1857--1858},
  year={2007},
  publisher={Oxford University Press}
}

@article{vaswani2017attention,
  title={Attention is all you need},
  author={Vaswani, Ashish and Shazeer, Noam and Parmar, Niki and Uszkoreit, Jakob and Jones, Llion and Gomez, Aidan N and Kaiser, {\L}ukasz and Polosukhin, Illia},
  journal={Advances in neural information processing systems},
  volume={30},
  year={2017}
}

@article{DBLP:journals/corr/abs-2006-11239,
  author       = {Jonathan Ho and
                  Ajay Jain and
                  Pieter Abbeel},
  title        = {Denoising Diffusion Probabilistic Models},
  journal      = {CoRR},
  volume       = {abs/2006.11239},
  year         = {2020},
  url          = {https://arxiv.org/abs/2006.11239},
  eprinttype    = {arXiv},
  eprint       = {2006.11239},
  bibsource    = {dblp computer science bibliography, https://dblp.org}
}

@article{samaniego2021ultrasensitive,
  title={Ultrasensitive molecular controllers for quasi-integral feedback},
  author={Samaniego, Christian Cuba and Franco, Elisa},
  journal={Cell Systems},
  volume={12},
  number={3},
  pages={272--288},
  year={2021},
  publisher={Elsevier}
}

@article{filo2024anti,
  title={Anti-windup strategies for biomolecular control systems facilitated by model reduction theory for sequestration networks},
  author={Filo, Maurice and Gupta, Ankit and Khammash, Mustafa},
  journal={Science Advances},
  volume={10},
  number={34},
  pages={eadl5439},
  year={2024},
  publisher={American Association for the Advancement of Science}
}

@article{qian2018realizing,
  title={Realizing ‘integral control’in living cells: how to overcome leaky integration due to dilution?},
  author={Qian, Yili and Del Vecchio, Domitilla},
  journal={Journal of The Royal Society Interface},
  volume={15},
  number={139},
  pages={20170902},
  year={2018},
  publisher={The Royal Society}
}

@article{qian2011neural,
  title={Neural network computation with {DNA} strand displacement cascades},
  author={Qian, Lulu and Winfree, Erik and Bruck, Jehoshua},
  journal={nature},
  volume={475},
  number={7356},
  pages={368--372},
  year={2011},
  publisher={Nature Publishing Group UK London}
}

@article{modi2021noise,
  title={Noise suppression in stochastic genetic circuits using PID controllers},
  author={Modi, Saurabh and Dey, Supravat and Singh, Abhyudai},
  journal={PLoS Computational Biology},
  volume={17},
  number={7},
  pages={e1009249},
  year={2021},
  publisher={Public Library of Science San Francisco, CA USA}
}

@article{hancock2022stabilization,
  title={Stabilization of antithetic control via molecular buffering},
  author={Hancock, Edward J and Oyarz{\'u}n, Diego A},
  journal={Journal of The Royal Society Interface},
  volume={19},
  number={188},
  pages={20210762},
  year={2022},
  publisher={The Royal Society}
}

@article{m2025multi,
  title={Multi-Layer Autocatalytic Feedback Enables Integral Control Amidst Resource Competition and Across Scales},
  author={M. Zand, Armin and Anastassov, Stanislav and Frei, Timothy and Khammash, Mustafa},
  journal={ACS Synthetic Biology},
  volume={14},
  number={4},
  pages={1041--1061},
  year={2025},
  publisher={ACS Publications}
}

@article{zand2024cascaded,
  title={Cascaded antithetic integral feedback for enhanced stability and performance},
  author={Zand, Armin M and Gupta, Ankit and Khammash, Mustafa},
  journal={IEEE Control Systems Letters},
  year={2024},
  publisher={IEEE}
}

@article{steel2019low,
  title={Low-burden biological feedback controllers for near-perfect adaptation},
  author={Steel, Harrison and Papachristodoulou, Antonis},
  journal={ACS synthetic biology},
  volume={8},
  number={10},
  pages={2212--2219},
  year={2019},
  publisher={ACS Publications}
}

@article{kelly2018synthetic,
  title={Synthetic negative feedback circuits using engineered small RNAs},
  author={Kelly, Ciar{\'a}n L and Harris, Andreas W K and Steel, Harrison and Hancock, Edward J and Heap, John T and Papachristodoulou, Antonis},
  journal={Nucleic acids research},
  volume={46},
  number={18},
  pages={9875--9889},
  year={2018},
  publisher={Oxford University Press}
}

@article{chang2026engineering,
  title={Engineering Cybergenetic Cell-Based Therapies},
  author={Chang, Ching-Hsiang and Arampatzis, Asterios and Balula, Samuel and Hou, Mucun and Filo, Maurice G and Chen, Mingzhe and Cella, Federica and Khammash, Mustafa},
  journal={bioRxiv},
  pages={2026--01},
  year={2026},
  publisher={Cold Spring Harbor Laboratory}
}

@article{agrawal2019vitro,
  title={In vitro implementation of robust gene regulation in a synthetic biomolecular integral controller},
  author={Agrawal, Deepak K and Marshall, Ryan and Noireaux, Vincent and Sontag, Eduardo D},
  journal={Nature communications},
  volume={10},
  number={1},
  pages={5760},
  year={2019},
  publisher={Nature Publishing Group UK London}
}

@article{zhang2024crispr,
  title={{CRISPR} perfect adaptation for robust control of cellular immune and apoptotic responses},
  author={Zhang, Yichi and Zhang, Shuyi},
  journal={Nucleic Acids Research},
  volume={52},
  number={16},
  pages={10005--10016},
  year={2024},
  publisher={Oxford University Press}
}

@article{salzano2025vivo,
  title={In Vivo Multicellular Feedback Control in Synthetic Microbial Consortia},
  author={Salzano, Davide and Shannon, Barbara and Grierson, Claire and Marucci, Lucia and Savery, Nigel J and di Bernardo, Mario},
  journal={ACS Synthetic Biology},
  year={2025},
  publisher={ACS Publications}
}

@article{martinelli2025multicellular,
  title={Multicellular PID control for robust regulation of biological processes},
  author={Martinelli, Vittoria and Fiore, Davide and Salzano, Davide and di Bernardo, Mario},
  journal={Journal of the Royal Society Interface},
  volume={22},
  number={222},
  pages={20240583},
  year={2025},
  publisher={The Royal Society}
}

@article{bier2015bmp,
  title={BMP gradients: A paradigm for morphogen-mediated developmental patterning},
  author={Bier, Ethan and De Robertis, Edward M},
  journal={Science},
  volume={348},
  number={6242},
  pages={aaa5838},
  year={2015},
  publisher={American Association for the Advancement of Science}
}

@article{lestas2010fundamental,
  title={Fundamental limits on the suppression of molecular fluctuations},
  author={Lestas, Ioannis and Vinnicombe, Glenn and Paulsson, Johan},
  journal={Nature},
  volume={467},
  number={7312},
  pages={174--178},
  year={2010},
  publisher={Nature Publishing Group UK London}
}

@article{mcadams1997stochastic,
  title={Stochastic mechanisms in gene expression},
  author={McAdams, Harley H and Arkin, Adam},
  journal={Proceedings of the National Academy of Sciences},
  volume={94},
  number={3},
  pages={814--819},
  year={1997},
  publisher={The National Academy of Sciences of the USA}
}

@article{otero2017chemical,
  title={Chemical reaction network theory elucidates sources of multistability in interferon signaling},
  author={Otero-Muras, Irene and Yordanov, Pencho and Stelling, Joerg},
  journal={PLoS computational biology},
  volume={13},
  number={4},
  pages={e1005454},
  year={2017},
  publisher={Public Library of Science San Francisco, CA USA}
}

@article{kim2006construction,
  title={Construction of an in vitro bistable circuit from synthetic transcriptional switches},
  author={Kim, Jongmin and White, Kristin S and Winfree, Erik},
  journal={Molecular systems biology},
  volume={2},
  number={1},
  pages={68},
  year={2006},
  publisher={John Wiley \& Sons, Ltd Chichester, UK}
}

@article{santos2020multistable,
  title={Multistable and dynamic {CRISPRi}-based synthetic circuits},
  author={Santos-Moreno, Javier and Tasiudi, Eve and Stelling, Joerg and Schaerli, Yolanda},
  journal={Nature communications},
  volume={11},
  number={1},
  pages={2746},
  year={2020},
  publisher={Nature Publishing Group UK London}
}

@article{wu2013engineering,
  title={Engineering of regulated stochastic cell fate determination},
  author={Wu, Min and Su, Ri-Qi and Li, Xiaohui and Ellis, Tom and Lai, Ying-Cheng and Wang, Xiao},
  journal={Proceedings of the National Academy of Sciences},
  volume={110},
  number={26},
  pages={10610--10615},
  year={2013},
  publisher={National Academy of Sciences}
}

@article{oyarzun2015design,
  title={Design of a bistable switch to control cellular uptake},
  author={Oyarz{\'u}n, Diego A and Chaves, Madalena},
  journal={Journal of The Royal Society Interface},
  volume={12},
  number={113},
  pages={20150618},
  year={2015},
  publisher={The Royal Society}
}

@article{siegal2009capacity,
  title={The capacity for multistability in small gene regulatory networks},
  author={Siegal-Gaskins, Dan and Grotewold, Erich and Smith, Gregory D},
  journal={BMC systems biology},
  volume={3},
  number={1},
  pages={96},
  year={2009},
  publisher={Springer}
}

@article{o2018homeostasis,
  title={Homeostasis, failure of homeostasis and degenerate ion channel regulation},
  author={O’Leary, Timothy},
  journal={Current Opinion in Physiology},
  volume={2},
  pages={129--138},
  year={2018},
  publisher={Elsevier}
}

@article{rai2026ultra,
  title={Ultra-high-throughput mapping of genetic design space},
  author={Rai, Kshitij and O’Connell, Ronan W and Piepergerdes, Trenton C and Wang, Yiduo and Brown, Lucas BC and Samra, Kian D and Wilson, Jack A and Lin, Shujian and Zhang, Thomas H and Ramos, Eduardo M and others},
  journal={Nature},
  pages={1--10},
  year={2026},
  publisher={Nature Publishing Group UK London}
}

@article{tatka2025speciated,
  title={Speciated evolution of oscillatory mass-action chemical reaction networks},
  author={Tatka, Lillian T and Smith, Lucian P and Sauro, Herbert M},
  journal={bioRxiv},
  year={2025}
}

@article{paladugu2006silico,
  title={In silico evolution of functional modules in biochemical networks},
  author={Paladugu, SR and Chickarmane, V and Deckard, A and Frumkin, JP and McCormack, M and Sauro, HM},
  journal={IEE Proceedings-Systems Biology},
  volume={153},
  number={4},
  pages={223--235},
  year={2006},
  publisher={IET}
}

@article{smart2024minimal,
  title={Minimal motifs for habituating systems},
  author={Smart, Matthew and Shvartsman, Stanislav Y and M{\"o}nnigmann, Martin},
  journal={Proceedings of the National Academy of Sciences},
  volume={121},
  number={41},
  pages={e2409330121},
  year={2024},
  publisher={National Academy of Sciences}
}

@article{eckert2024biochemically,
  title={Biochemically plausible models of habituation for single-cell learning},
  author={Eckert, Lina and Vidal-Saez, Maria Sol and Zhao, Ziyuan and Garcia-Ojalvo, Jordi and Martinez-Corral, Rosa and Gunawardena, Jeremy},
  journal={Current Biology},
  volume={34},
  number={24},
  pages={5646--5658},
  year={2024},
  publisher={Elsevier}
}

@article{kim2012mechanism,
  title={A mechanism for robust circadian timekeeping via stoichiometric balance},
  author={Kim, Jae Kyoung and Forger, Daniel B},
  journal={Molecular systems biology},
  volume={8},
  pages={630},
  year={2012}
}

@article{jeong2022combined,
  title={Combined multiple transcriptional repression mechanisms generate ultrasensitivity and oscillations},
  author={Jeong, Eui Min and Song, Yun Min and Kim, Jae Kyoung},
  journal={Interface Focus},
  volume={12},
  number={3},
  year={2022},
  publisher={The Royal Society}
}

\newpage
\section*{Methods}\phantomsection
\edef\@currentlabelname{Methods}\label{sec:methods}

\subsection*{Components of the RL Loop}
We begin by defining the agent and its action space (reaction library), and then proceed to the environment and its state space (I/O CRNs). We then describe the translator that maps CRNs into an agent-interpretable representation, followed by the policy architecture used by the agent.

\paragraph{The agent and action space.}
The agent is responsible for proposing the next reaction to append to the current incomplete I/O CRN, iterating until the I/O CRN is complete. We formalize the agent's decision as sampling an action from a reaction library, conditioned on the current translated CRN representation. Equivalently, the agent defines a conditional distribution over the next reaction given the current CRN state, i.e.,
$\mathbb P\!\left(\text{next reaction}\mid \text{CRN}\right)$, as schematized in Fig.~\ref{Fig:Method}A.

Informally, the action space $\mathcal{A}$ corresponds to the set of reactions that may appear in a valid CRN. Because different application settings permit different classes of reactions, we treat $\mathcal{A}$ as a user- or task-chosen \emph{reaction library}, which is a core component of our formulation containing all permissible reaction \templates. Let reaction \templates in the library be indexed by $\ell\in\{1,\dots,M\}$, where each \template $\ell$ corresponds to a human-interpretable reaction template (stoichiometry together with associated kinetic parameter variables). When external inputs are present, each reaction may additionally be configured with an \emph{input-influence pattern} indicating which input channels modulate its propensity. Concretely, let there be $n_u$ input channels and let $\mathcal{G}$ denote the (finite) set of admissible input-influence patterns (e.g., $\mathcal{G}\subseteq\{0,1\}^{n_u}$, where a binary vector $g\in\mathcal{G}$ specifies which inputs act on the reaction). Reaction \template $\ell$ has $p_\ell\in\mathbb{N}_0$ parameters with domains $\mathbb{S}_{\ell,\ell'}$ for $\ell'=1,\dots,p_\ell$. An action is then a tuple consisting of (i) a discrete reaction-type index, (ii) its continuous parameter vector, and (iii) an input-influence choice:
\begin{equation}
\label{eq:action-space}
\begin{aligned}
\mathcal{A}
\;\defeq\;
\left\{
(\ell,\theta,g)\ \middle|\ \right. 
& \ell \in\{1,\dots,M\}\\
& \theta =(\theta_1,\dots,\theta_{p_\ell}),\ 
\theta_{\ell'} \in\mathbb{S}_{\ell, \ell'}\ \forall \ell',\ 
g\in\mathcal{G} \left.
\right\}.
\end{aligned}
\end{equation}
When a task requires no external input assignments, we take $n_u=0$ and $\mathcal{G}=\{\emptyset\}$, reducing to the $(\ell,\theta)$ action form. Note that $\mathcal{A}$ is a hybrid space with both discrete (reaction template index) and continuous (kinetic parameters) components.

\paragraph{The environment and state space.}
Let $s_t$ denote the environment state at construction step $t$, defined as the current (possibly incomplete) I/O CRN obtained by starting from the user-provided \startercrn\ and appending $t$ reactions. Thus, $s_0$ is exactly the \startercrn, and for $t>0$ the state $s_t$ encodes the set of selected reaction \templates together with their chosen parameter values and input-influence assignments.

Formally, the state space is the combinatorial set of all such CRNs reachable by appending up to $m$ reactions from the library: 
\begin{equation} 
\begin{aligned} 
\mathcal{S} \defeq \bigcup_{t=0}^{m-1} \mathcal{S}_t, ~~ \text{with} ~~ \mathcal S_t \defeq \binom{\mathcal{A}}{t} \defeq \{\, \mathcal B \subseteq \mathcal{A} \mid |\mathcal B| = t \,\}. 
\end{aligned} 
\end{equation} 

We detail the actual size of the search space under mass action kinetics in a subsequent section.
In practice, we enforce validity constraints during construction (e.g., preventing duplicate reaction types and considering the \startercrn).
We next describe how a state $s_t$ is mapped into the agent-interpretable representation on which the policy conditions.

\paragraph{The translator.}
The translator maps a human-readable CRN state $s_t$ (a reaction list with parameters and input influences) into a fixed-format, agent-interpretable observation. Concretely, it encodes, for each library reaction \template, whether it is present in the current I/O CRN and, if so, its current parameter value(s) and input-influence assignment. This yields a structured representation of constant dimension across all $t$, enabling the policy to operate on incomplete CRNs of varying sizes. A running example is shown in Fig.~\ref{Fig:Method}A (left), where a toy I/O CRN is translated into this fixed-format encoding using an example library; the reaction highlighted by the red box illustrates how a single library entry is represented.

\paragraph{Policy implementation.}
For compactness, we describe the policy architecture for the case where input assignment is not required (Fig.~\ref{Fig:Method}B). The agent is parameterized by neural-network weights $\phi$ and defines a conditional distribution over the next reaction type and its kinetic parameters given the current translated CRN state $s$, where we drop the subscript $t$ for notational convenience. As illustrated in the left panel of Fig.~\ref{Fig:Method}B and zoomed-in on the right, the policy is implemented with a shared network trunk that produces an embedding of $s$, followed by two heads, and is written autoregressively as
\begin{equation}
\label{eq:policy-factorization}
\pi_\phi(\ell,\theta \mid s)
\;=\;
\pi_\phi(\ell \mid s)\,
\pi_\phi(\theta \mid s,\ell),
\end{equation}
where we recall that $\ell$ denotes the reaction \template index (or ID) in the library, and $\theta$ denotes its parameter values. 
The \emph{structure head} outputs logits over reaction \templates $\ell\in\{1,\dots,M\}$; logits for reaction \templates already present in the current partial CRN are masked to enforce uniqueness, yielding the categorical distribution $\pi_\phi(\ell\mid s_t)$. Conditioned on the sampled \template $\ell$, the \emph{parameter head} outputs a continuous distribution over $\theta$ (we use a lognormal family), yielding $\pi_\phi(\theta\mid s,\ell)$. 
During sampling, the agent draws $\ell\sim \pi_\phi(\ell\mid s)$ and then $\theta\sim \pi_\phi(\theta\mid s,\ell)$. The corresponding log-probabilities are summed up in log-space.
Furthermore, we compute an weighted entropy regularizer by combining the Shannon entropy of the discrete component and the differential entropy of the continuous component:
\begin{equation}
\label{eq:entropy-decomposition}
\mathcal{H}_{\pi_\phi}^w(s)
\;=\;
\mathcal{H}\!\left[\pi_\phi(\cdot\mid s)\right]
\;+\;
w\,\mathbb{E}_{\ell\sim \pi_\phi(\cdot\mid s)}
\Big[\,
\mathcal{H}\!\left[\pi_\phi(\cdot \mid s,\ell)\right]
\,\Big],
\end{equation}
where the expectation appears because the parameter distribution $\pi_\phi(\theta\mid s,\ell)$ (and thus its differential entropy) depends on the reaction \template $\ell$, so we average over $\ell$ under $\pi_\phi(\ell\mid s)$ to obtain a single entropy value for the joint policy at state $s$. Finally, $w$ is a training hyperparameter controlling the relative strength of the continuous-entropy term, which we discuss in the subsequent training section.


\subsection*{The Training}
Training proceeds via episodic construction rollouts of fixed length $m$. Starting from the initial state $s_0$ (the \startercrn), the agent repeatedly samples an action and the environment deterministically applies it through the state transition operator $\mathcal T$ (by appending the proposed reaction):
\begin{equation}
\label{eq:rollout}
a_t \sim \pi_\phi(\cdot\mid s_t),\qquad s_{t+1}=\mathcal{T}(s_t,a_t),\qquad t=0,1,\dots,m-1,
\end{equation}
yielding a trajectory $\tau=(s_0,a_0,\dots,s_{m-1},a_{m-1},s_m)$, where $s_m$ is a \emph{complete} I/O CRN. The completed CRN $s_m$ is then simulated and evaluated using a task-specific objective to yield a scalar loss $\mathcal{L}_{\mathrm{task}}(s_m)$ (defined by the user for each application, as illustrated in our examples). Because partial CRNs can exhibit behavior that is not predictive of the final circuit, we assign reward only at termination and optimize \emph{terminal} performance.

\paragraph{Policy optimization.}
We define the terminal reward as the negative task loss,
\begin{equation}
\label{eq:terminal-reward}
\mathcal{R}(\tau)\;\defeq\;\mathcal{R}(s_m)\;\defeq\;-\mathcal{L}_{\mathrm{task}}(s_m).
\end{equation}
Under the standard REINFORCE formulation \cite{williams1992simple} with terminal rewards, the policy is trained by solving the expected-return maximization problem
\begin{equation}
\label{eq:reinforce-objective}
\max_{\phi}\;
\mathbb{E}_{\tau\sim p_\phi}\!\left[\mathcal{R}(\tau)\right]
\;=\;
\mathbb{E}_{\tau\sim p_\phi}\!\left[-\mathcal{L}_{\mathrm{task}}(s_m)\right],
\end{equation}
where $p_\phi(\tau)$ denotes the rollout distribution induced by the policy $\pi_\phi$ and the deterministic transition $\mathcal{T}$. In \crngen, we build on this objective and introduce four modifications—(i) a risky (top-$K$) objective that emphasizes high-performing rollouts \cite{chow2014algorithms}, (ii) baseline subtraction using the worst reward among the top-$K$ samples, (iii) hybrid entropy regularization combining discrete (Shannon) and continuous (differential) entropy terms \cite{delalleau2019discrete, geist2019theory}, and (iv) a self-imitation learning (SIL) term based on a hall-of-fame replay buffer \cite{oh2018self} —each described in detail in what follows.

\paragraph{Risk-sensitive (top-$K$) objective with baseline subtraction.}
Standard REINFORCE maximizes the expected return, which in batch form averages rewards across sampled rollouts. To reduce this averaging effect and emphasize rare high-performing solutions, we skew the objective toward the \emph{upper tail} of the reward distribution. Let $q_\alpha$ denote the $\alpha$-quantile of the terminal reward $\mathcal R(\tau)$ under $\tau\sim p_\phi$ (i.e., $\mathbb{P}(\mathcal R(\tau)\le q_\alpha)=\alpha$). The corresponding upper-tail (top-quantile) objective is
\[
J_\alpha(\phi)\;\defeq\;\mathbb{E}\!\left[\,\mathcal R(\tau)\ \middle|\ \mathcal R(\tau)\ge q_\alpha\,\right],
\]
which is closely related to CVaR-style risk-sensitive optimization and the \emph{risky policy gradients} viewpoint \cite{chow2014algorithms}. In practice, we estimate this objective from a batch of $N$ rollouts $\{\tau^{(i)}\}_{i=1}^N$ with terminal rewards $r_i\defeq \mathcal{R}(\tau^{(i)})$. Let $r_{(1)}\ge \cdots \ge r_{(N)}$ be the sorted rewards and choose $K=\lceil N(1-p)\rceil$, so that the empirical threshold $r_{(K)}$ estimates $q_\alpha$. We further apply baseline subtraction with the batch baseline defined as the \emph{worst} reward among the top-$K$ samples,
\[
\mathcal{B}_{\mathrm{top}K}\;\defeq\; r_{(K)}.
\]
The resulting baseline-shifted top-$K$ objective is
\begin{equation}
\label{eq:topk-baseline}
\begin{aligned}
J_{\mathrm{RL}}
&\;\defeq\;
\frac{1}{K}\sum_{i=1}^{N}\big(r_i-\mathcal{B}_{\mathrm{top}K}\big)\,
\mathbf{1}\!\left\{r_i \ge \mathcal{B}_{\mathrm{top}K}\right\} \\
& \;=\;
\frac{1}{K}\sum_{i=1}^{K}\big(r_{(i)}-\mathcal{B}_{\mathrm{top}K}\big),
\end{aligned}
\end{equation}
so that only rollouts in the empirical top quantile contribute, with non-negative shifted returns, while all remaining rollouts are excluded from this objective.

\paragraph{Entropy bonus.}
To encourage exploration in the hybrid action space, we add an entropy bonus computed along construction steps. Using the per-state hybrid policy entropy $\mathcal{H}_{\pi_\phi}^w(s)$ defined in \eqref{eq:entropy-decomposition}, we form
\begin{equation}
\label{eq:entropy-trajectory}
J_{H}^w
\;\defeq\;
\mathbb{E}_{\tau}\!\left[\sum_{t=0}^{m}\mathcal{H}_{\pi_\phi}^w(s_t)\right],
\end{equation}
and weight it with a training coefficient $\lambda_H\ge 0$. 
Note that the gradient of this entropy regularization term is implemented using a pathwise stop-gradient approximation. This omits the score-function cross term yielding a lower-variance (but biased) entropy-gradient estimator.

\paragraph{Self-imitation learning (SIL).}
To prevent ``forgetting'' rare but valuable solutions, we maintain a hall-of-fame replay buffer containing the best complete CRNs observed so far. Let $S_k$ denote the set of complete CRNs sampled in episode $k$, and define the cumulative sample set up to episode $k$ as $S_{\le k}=\bigcup_{\bar k =1}^{k} S_{\bar k}$. For a hall-of-fame capacity $k_{\mathrm{HOF}}$, we define
\begin{equation}
\label{eq:hof}
\mathrm{HOF}_k \;\defeq\; \operatorname{TopK}_{k_{\mathrm{HOF}}}\!\big(S_{\le k}; \mathcal R\big),
\end{equation}
i.e., the set of the $k_{\mathrm{HOF}}$ highest-reward complete CRNs observed so far. Within the current batch, let
\begin{equation}
\label{eq:batch-best}
R_m^{*} \;\defeq\; \max_{s\in S_k} \mathcal R (s)
\end{equation}
be the best reward achieved in the current iteration. We then add a SIL objective that increases the likelihood of sampling hall-of-fame solutions only when they outperform the current batch best:
\begin{equation}
\label{eq:sil}
\begin{aligned}
&J_{\mathrm{SIL}}(\phi)
\;\defeq\;
-\;\mathbb{E}_{s\sim \mathrm{HOF}_k}\Big[\,(\mathcal R(s)-R_t^{*})_{+}\;\log \bar p_\phi(s)\Big], \\
& \qquad \qquad \text{with} \quad (x)_{+}\defeq \max(0,x),
\end{aligned}
\end{equation}
where $\bar p_\phi(s)$ denotes the probability (under the current policy) of generating the complete CRN $s$, equivalently the likelihood of the action sequence that constructs $s$, with per-step log-probabilities summed in log-space as described above.

\paragraph{Final objective.}
We combine the risk-sensitive RL term, the entropy bonus, and SIL:
\begin{equation}
\label{eq:final-objective}
J(\phi)
\;\defeq\;
J_{\mathrm{RL}}(\phi)
\;+\;
\lambda_H\,J_H^w(\phi)
\;+\;
\lambda_{\mathrm{SIL}}\,J_{\mathrm{SIL}}(\phi),
\end{equation}
and optimize the agent by minimizing the corresponding loss
$\mathcal{L}(\phi)\defeq -J(\phi)$. The hyperparameters are $(p,k,w, \lambda_H,\lambda_{\mathrm{SIL}},k_{\mathrm{HOF}})$.

\subsection*{Task-dependent loss functions}
Let $\net$ denote an input--output chemical reaction network. Given an input signal $u$ and initial condition $x_0$, we denote the (vector-valued) network output trajectory by $y(t; x_0, u)$; we omit $x_0$ and/or $u$ from the notation when they are fixed by the task. Each task defines a loss functional $\mathcal{L}_{\text{task}}(\net)$ that quantifies how well $\net$ matches a desired behavior. Unless stated otherwise, we measure output mismatch with an $\ell_1$ error
\[
e(t)\;\defeq\;\|y(t; x_0, u)-r\|_1,
\]
where $r$ denotes a desired reference (e.g., a setpoint or target trajectory). When performance must hold across multiple tested conditions (inputs, disturbances, and/or initial conditions), we aggregate by averaging over the corresponding evaluation set.

For time-integrated objectives, we use a weighted integral operator
\begin{align}
\mathcal{I}[e]
&\defeq \int_{0}^{t_{\mathrm{f}}} w_d(t)\, e(t)\,dt, \\
w_d(t)
&=
\begin{cases}
0.25, & 0 \le t < t_{\mathrm{f}}/5,\\
1, & t_{\mathrm{f}}/5 \le t < 4t_{\mathrm{f}}/5,\\
2, & 4t_{\mathrm{f}}/5 \le t \le t_{\mathrm{f}},
\end{cases}
\label{eq:weighted_integral}
\end{align}
where $t_{\mathrm{f}}$ is the final simulation time and $w_d(t)$ emphasizes late-time accuracy (steady-state performance) while still penalizing transient error. Next, we specify the task-specific losses used in this work.

\paragraph{Dose--response shaping.}
For this task we fix the initial condition to $x_0=0$. For each tested input level $u\in\mathcal{U}$, we define the input-conditional loss
\begin{equation}
\mathcal{L}_{\mathrm{DR}}^{u}(\net)
\;\defeq\;
\mathcal{I}\!\big[\|y(t;u)-r(u)\|_1\big],
\end{equation}
where $r(u)$ is the target dose--response function. The overall dose--response loss averages across the evaluated input set:
\begin{equation}
\mathcal{L}_{\mathrm{DR}}(\net)
\;\defeq\;
\frac{1}{|\mathcal{U}|}\sum_{u\in\mathcal{U}}\mathcal{L}_{\mathrm{DR}}^{u}(\net).
\end{equation}

\paragraph{Robust perfect adaptation.}
For this task, we evaluate performance across setpoints $u_1\in\mathcal{U}_1$ and disturbances $u_2\in\mathcal{U}_2$, with reference $r(u_1)$ independent of $u_2$. We define the overall RPA loss by averaging the time-integrated tracking error over all tested $(u_1,u_2)$ conditions:
\begin{equation}
\mathcal{L}_{\mathrm{RPA}}(\net)
\;\defeq\;
\frac{1}{|\mathcal{U}_1|\,|\mathcal{U}_2|}
\sum_{u_1\in\mathcal{U}_1}\ \sum_{u_2\in\mathcal{U}_2}
\mathcal{I}\!\left[\|y(t;u_1,u_2)-r(u_1)\|_1\right].
\end{equation}

\paragraph{Fate decision circuits.}
For this task, we evaluate fate decisions across a set of initial conditions $x_0\in\mathcal{X}_0$, where each $x_0$ is assigned a basin-specific reference trajectory $r(x_0)$. We define the overall loss by averaging the time-integrated tracking error over all tested initial conditions:
\begin{equation}
\mathcal{L}_{\mathrm{Fate}}(\net)
\;\defeq\;
\frac{1}{|\mathcal{X}_0|}
\sum_{x_0\in\mathcal{X}_0}
\mathcal{I}\!\left[\|y(t;x_0)-r(x_0)\|_1\right].
\end{equation}

\paragraph{Logic circuits.}
For each truth-table condition $c\in\mathcal{C}_{\mathrm{LOGIC}}$, we measure terminal-time disagreement at $t_{\mathrm{f}}$ by
\begin{equation}
\mathcal{L}_{\mathrm{LOGIC}}^{c}(\net)
\;\defeq\;
\left\| y(t_{\mathrm{f}};c)-r(c)\right\|_1.
\end{equation}
We then average across all truth-table conditions:
\begin{equation}
\mathcal{L}_{\mathrm{LOGIC}}(\net)
\;\defeq\;
\frac{1}{|\mathcal{C}_{\mathrm{LOGIC}}|}
\sum_{c\in\mathcal{C}_{\mathrm{LOGIC}}}
\mathcal{L}_{\mathrm{LOGIC}}^{c}(\net).
\end{equation}

\paragraph{Oscillators.}
To promote sustained periodic behavior we maximize a \emph{periodicity index} computed from the normalized autocorrelation of the centered output. For a scalar output $y(t)$, define the time-averaged mean
\[
\bar{y}\;\defeq\;\frac{1}{t_{\mathrm{f}}}\int_{0}^{t_{\mathrm{f}}} y(t)\,dt,
\]
the centered signal
\begin{align}
y_c(t) &\triangleq y(t) - \bar{y},
\end{align}
and the normalized autocorrelation (for lags $\tau\ge 0$)
\begin{align}
R_y(\tau)
&\triangleq\frac{\int_{0}^{t_{\mathrm{f}}-\tau} \bar y(t)\,\bar y(t+\tau)\,dt}{\int_{0}^{t_{\mathrm{f}}} \bar y^2(t)\,dt}.
\end{align}
We then define the periodicity index as the height of the first positive-lag local maximum of $R_y$,
\begin{align}
\rho_1[y]
&\defeq\max_{\tau>0}\Bigl\{R_y(\tau)\ \big|\ R_y'(\tau)=0,\ R_y''(\tau)<0\Bigr\}.
\end{align}
In addition to periodicity, we constrain the oscillation \emph{center} (time average) to match a prescribed target $\mu^\star\in\mathbb{R}$. We thus define the oscillator loss
\begin{align}
\mathcal{L}_{\mathrm{Osc}}(\net)
\;\defeq\;
\bigl(1-\rho_1\!\left[y\right]\bigr)
\;+\;
\lambda_\mu\,
\bigl|\bar{y}-\mu^\star\bigr|,
\end{align}
where $\lambda_\mu\ge 0$ weights the center-matching term.

\paragraph{Oscillators with frequency tuning.}
For frequency-tunable oscillators we include terms that (i) match a target frequency and (ii) discourage amplitude damping.

We estimate the oscillation period by averaging successive inter-peak intervals. Let $t_i$ denote successive peak times of $|y_c(t)|$, let $n_p$ be the number of detected peaks in $[0,t_{\mathrm{f}}]$, and define
\begin{align}
T[y]
&\defeq \frac{1}{n_p-1}\sum_{i=1}^{n_p-1}\bigl(t_{i+1}-t_i\bigr).
\end{align}
To quantify damping, define peak amplitudes $A_i\defeq|y_c(t_i)|$ and the damping index
\begin{align}
\zeta[y]
&\defeq \frac{1}{n_p-1}\sum_{i=1}^{n_p-1}\frac{A_i}{A_{i+1}}.
\end{align}
For a single evaluated condition $u\in\mathcal{U}$ (where $u$ specifies the target frequency), we define the per-condition loss
\begin{equation}
\begin{aligned}
\mathcal{L}_{\mathrm{OscF}}^{u}(\net)
\;\defeq\;
\bigl(1-\rho_1[y(\cdot;u)]\bigr)
\;&+\;
\lambda_f\left|\frac{1}{T[y(\cdot;u)]}-u\right| \\
\;&+\;
\lambda_d\bigl(\zeta[y(\cdot;u)]-1\bigr),
\end{aligned}
\end{equation}
where $\rho_1[\cdot]$ is the periodicity index defined above, and $\lambda_f,\lambda_d\ge 0$ are weights. The overall loss averages across the evaluated conditions:
\begin{equation}
\mathcal{L}_{\mathrm{OscF}}(\net)
\;\defeq\;
\frac{1}{|\mathcal{U}|}\sum_{u\in\mathcal{U}}\mathcal{L}_{\mathrm{OscF}}^{u}(\net).
\end{equation}

\paragraph{Habituation and sensitization losses.}
In this application, we target two adaptive response classes---habituation and sensitization \cite{smart2024minimal,eckert2024biochemically}---using a common loss construction based on three qualitative hallmarks: (i) a systematic change in response amplitude across repeated pulses within a pulse train (decreasing for habituation, increasing for sensitization), (ii) recovery of the output toward its pre-stimulus fixed point after a sufficiently long OFF gap (\emph{spontaneous recovery}), and (iii) a stronger cumulative effect under higher-frequency stimulation (\emph{frequency sensitivity}). The two tasks differ only in the direction of the within-train peak progression; all other terms are shared.

We evaluate these hallmarks over input levels $u\in\mathcal{U}$ and pulse shapes $\mathbf{p}\in\mathcal{P}$, where each pulse shape $\mathbf{p}=(t_{\mathrm{on}},t_{\mathrm{off}})$ defines both a stimulation frequency and a duty cycle. Each condition $(u,\mathbf{p})$ consists of three phases: a pre-gap pulse train, an OFF gap of duration $t_{\mathrm{gap}}$, and a post-gap pulse train (see Extended Data Fig.~\ref{EDF:Habituation}B,C, red pulses). Let $y(t;u,\mathbf{p})$ denote the output trajectory, and let $a_k^{\mathrm{pre}}(u,\mathbf{p})$ and $a_k^{\mathrm{post}}(u,\mathbf{p})$ denote the detected peak amplitudes (measured relative to the pre-stimulus fixed point) during the pre-gap and post-gap pulse trains, respectively.

To treat habituation and sensitization in a unified way, we define the sign
\begin{equation}
s \;\defeq\;
\begin{cases}
+1, & \text{habituation},\\
-1, & \text{sensitization},
\end{cases}
\end{equation}
so that the desired trend is always encoded by
\[
s\,\bigl(a_{k+1}-a_k\bigr)\le 0,
\]
that is, decreasing peaks for habituation and increasing peaks for sensitization.

For a fixed condition $(u,\mathbf{p})$, let $n_{\mathrm{pre}}$ be the number of detected peaks in the pre-gap pulse train, and define the signed consecutive-peak ratios
\begin{equation}
r_k(u,\mathbf{p})
\;\defeq\;
\left(\frac{a_{k+1}^{\mathrm{pre}}(u,\mathbf{p})}{a_k^{\mathrm{pre}}(u,\mathbf{p})+\varepsilon}\right)^s,
\qquad k=1,\dots,n_{\mathrm{pre}}-1,
\end{equation}
where $\varepsilon>0$ is a small constant for numerical stability. We quantify the within-train trend by the log-median of these ratios:
\begin{equation}
\mathcal{L}_{\mathrm{trend}}^{u,\mathbf{p}}(\net)
\;\defeq\;
\log\!\Bigl(\operatorname{median}\{r_k(u,\mathbf{p})\}_{k=1}^{n_{\mathrm{pre}}-1}+\varepsilon\Bigr).
\end{equation}
This term is small when the pre-gap pulse amplitudes evolve in the desired direction.

To enforce spontaneous recovery, let $t_{\mathrm{gap}}^{-}$ denote the end of the OFF gap and let $y_{\mathrm{fp}}(u)$ be the pre-stimulus fixed-point output under zero input. We define
\begin{align*}
\mathcal{L}_{\mathrm{gap}}^{u,\mathbf{p}}(\net)
&\;\defeq\;
\bigl|y(t_{\mathrm{gap}}^{-};u,\mathbf{p})-y_{\mathrm{fp}}(u)\bigr| \\
&\quad+\;
\Bigl[a_1^{\mathrm{pre}}(u,\mathbf{p})-a_1^{\mathrm{post}}(u,\mathbf{p})\Bigr]_+,
\end{align*}
which jointly encourages the output to relax back toward its original fixed point during the gap and to recover a comparable first-pulse response after the gap.

To reward stronger adaptation under higher-frequency stimulation, we define the signed adaptation score
\begin{equation}
\Delta(u,\mathbf{p})
\;\defeq\;
-\;s\,
\frac{a_{n_{\mathrm{pre}}}^{\mathrm{pre}}(u,\mathbf{p})-a_1^{\mathrm{pre}}(u,\mathbf{p})}
{a_1^{\mathrm{pre}}(u,\mathbf{p})+\varepsilon},
\end{equation}
so that larger $\Delta(u,\mathbf{p})$ always corresponds to a stronger effect, whether habituation (larger decrease) or sensitization (larger increase). Ordering pulse shapes by increasing frequency as $\mathbf{p}_1,\dots,\mathbf{p}_{|\mathcal P|}$, we penalize violations of monotonic strengthening:
\begin{equation}
\mathcal{L}_{\mathrm{freq}}^{u}(\net)
\;\defeq\;
\frac{1}{|\mathcal{P}|-1}\sum_{j=1}^{|\mathcal{P}|-1}
\bigl[\Delta(u,\mathbf{p}_j)-\Delta(u,\mathbf{p}_{j+1})\bigr]_+,
\end{equation}
with $\mathcal{L}_{\mathrm{freq}}^{u}(\net)\equiv 0$ when $|\mathcal P|=1$.

The per-condition loss combines the within-train trend and spontaneous-recovery terms:
\begin{equation}
\mathcal{L}_{\mathrm{Hab/Sens}}^{u,\mathbf{p}}(\net)
\;\defeq\;
\mathcal{L}_{\mathrm{trend}}^{u,\mathbf{p}}(\net)
\;+\;
\lambda_{\mathrm{gap}}\mathcal{L}_{\mathrm{gap}}^{u,\mathbf{p}}(\net),
\end{equation}
and the overall loss averages across inputs and pulse shapes, with an additional frequency-sensitivity penalty:
\begin{align*}
\mathcal{L}_{\mathrm{Hab/Sens}}(\net)
&\;\defeq\;
\frac{1}{|\mathcal U|\,|\mathcal P|}
\sum_{u\in\mathcal U}\sum_{\mathbf{p}\in\mathcal P}
\mathcal{L}_{\mathrm{Hab/Sens}}^{u,\mathbf{p}}(\net) \\
&\quad+\;
\lambda_{\mathrm{freq}}\,
\frac{1}{|\mathcal U|}\sum_{u\in\mathcal U}\mathcal{L}_{\mathrm{freq}}^{u}(\net),
\end{align*}
where $\lambda_{\mathrm{gap}},\lambda_{\mathrm{freq}}\ge 0$ are weighting coefficients.

\paragraph{Stochastic objectives (mean and noise).}
For stochastic CRNs we optimize moments of the output. Under an evaluation condition $c\in\mathcal{C}$, let
\[
\mu(t;c)\;\defeq\;\mathbb{E}[y(t;c)]
\]
denote the output mean. We define the mean-tracking loss
\begin{equation}
\mathcal{L}_{\mathrm{MEAN}}^{c}(\net)
\;\defeq\;
\mathcal{I}\!\left[\|\mu(t;c)-r(c)\|_1\right].
\end{equation}
To penalize variability, we use the (component-wise) coefficient of variation
\[
\mathrm{CV}(t;c)\;\defeq\;\frac{\sigma(t;c)}{\mu(t;c)+\varepsilon},
\]
where $\sigma(t;c)$ is the output standard deviation and $\varepsilon>0$ avoids division by zero. The variability loss is
\begin{equation}
\mathcal{L}_{\mathrm{CV}}^{c}(\net)
\;\defeq\;
\mathcal{I}\!\left[\|\mathrm{CV}(t;c)\|_1\right].
\end{equation}
We combine the two objectives as
\begin{equation}
\mathcal{L}_{\mathrm{MEAN+CV}}^{c}(\net)
\;\defeq\;
\mathcal{L}_{\mathrm{MEAN}}^{c}(\net)
+\lambda_{\mathrm{CV}}\mathcal{L}_{\mathrm{CV}}^{c}(\net),
\end{equation}
with $\lambda_{\mathrm{CV}}\ge 0$. The overall stochastic loss averages across evaluated conditions:
\begin{equation}
\mathcal{L}_{\mathrm{MEAN+CV}}(\net)
\;\defeq\;
\frac{1}{|\mathcal{C}|}\sum_{c\in\mathcal{C}}\mathcal{L}_{\mathrm{MEAN+CV}}^{c}(\net),
\end{equation}
(or, when conditions are grouped, by averaging within groups and then across groups).

\subsection*{Size of the search space}
\label{sec:sizeofthesearchspace}
Candidate topologies scale combinatorially with the reaction-template library. Let $\mathbb{L}$ be the library of admissible reaction templates and let $\mathbb{L}_0\subseteq\mathbb{L}$ denote templates fixed a priori (e.g., those in the \startercrn) or excluded from sampling. The effective library is $\mathbb{L}^-=\mathbb{L}\setminus\mathbb{L}_0$. Let $M \defeq |\mathbb L^-|$ be the size of the effective library. Appending $m$ reactions corresponds to selecting a size-$m$ subset of $\mathbb{L}^-$, so the number of topologically distinct candidates is
\begin{align}
n_{\mathrm{CRN}}(m, M) &= \binom{M}{m}.
\end{align}

The library size $M$ depends on the reaction family and the number of species.

\paragraph{Mass-action kinetics.}
Under mass-action kinetics, a template reaction is specified by a reactant and a product complex (including the null complex). We say that the reaction is of doubly order $o$ if the reactant and product complexes contain at most $o$ molecules drawn from a species set of size $n$. Therefore, given $n$ and $o$, the number of all possible complexes is
\begin{align}
n_c(n, o)
&= \sum_{i=0}^{o}\mathrm{MultiSet}(n,i),
\end{align}
where 
\begin{align}
\mathrm{MultiSet}(a,b) \defeq \binom{a+b-1}{b}
\end{align}
denote the number of multisets of cardinality $b$ drawn from a set of size $a$.
It can be shown that the sum can be simplified to yield
\begin{equation}
n_c(n, o) = \binom{n+o}{n}.
\end{equation}
Finally, pairing distinct reactant/product complexes yields
\begin{align}
M
&= n_c(n,o) \big[ n_c(n, o) -1 \big] + 1,
\end{align}
where the final term accounts for the null reaction (empty-to-empty).

For instance, for $n=3$ species, order $o=2$, and $m=5$ reactions, the library size is $M=91$, and the number of possible CRN topologies is $n_{\text{CRN}} = 46,504,458$. Of course this excludes the rate parameters which further expands the space. Adding one more reaction, that is $m=6$, without changing the number of species and reaction order increases the possibilities by over an order of magnitude to $n_{\text{CRN}} = 666,563,898$. Allowing $n=4$ species increases the library size to $M=211$ and the number of CRN topologies to $n_{\text{CRN}} = 114,081,819,852$.

\paragraph{Hill-type kinetics.}
For Hill-type (production-only) templates, the number of distinct product complexes of size up to $n_{\mathrm{prod}}$ is 
$$\sum_{n_i=1}^{n_{\mathrm{prod}}}\mathrm{MultiSet}(n,n_i).$$
Signed regulator (activator or repressor enzyme) sets of size up to $n_{\mathrm{REG}}$ contribute a factor
\begin{align}
\sum_{n_i=1}^{n_{\mathrm{REG}}}\binom{n}{n_i}\,2^{n_i},
\end{align}
so the Hill library size is
\begin{align}
M
&= \sum_{n_i=1}^{n_{\mathrm{prod}}} \mathrm{MultiSet}(n,n_i)
\sum_{n_i=1}^{n_{\mathrm{REG}}}\binom{n}{n_i}\,2^{n_i}.
\end{align}
For example, allowing only on product per reaction ($n_{\text{prod}} = 1$) and up to two regulators ($n_{\text{REG}} = 2$), we have $M=432$ reactions.

\subsection*{Implementation and computational resources}
The \crngen\ implementation will be made fully available on GitHub with the final version of this preprint. The software is written in Python, using PyTorch for the reinforcement-learning loop and PyCUDA for GPU-accelerated stochastic simulation.


Deterministic analyses other than the dose-response experiments were run on a workstation (WS1) equipped with an AMD Ryzen Threadripper PRO 7985WX CPU and an NVIDIA RTX 4500 Ada GPU, with computations distributed across 64 CPU cores. Stochastic simulations, which are predominantly GPU-bound, were performed on a machine (WS2) equipped with an Intel Xeon W-2123 CPU and NVIDIA Titan RTX GPUs; these resources were provided by the D-BSSE ETH Zurich. Dose-response experiments were conducted on a computing cluster (CLS) with Intel Xeon Gold 6330 CPUs using 40 CPU cores and with Nvidia A800 GPUs, provided by the Academy of Mathematics and Systems Science, Chinese Academy of Sciences.

Deterministic simulations for logic-circuit training runs were executed on {WS1}. For the logic-circuit experiments (5 species, 6 reactions), training with batch size $1024$ for $300$ epochs required approximately {5 hours}. Stochastic simulations were executed on {WS2}: for all SSA-based evaluations we used {1000 SSA simulations per batch element}. The stochastic RPA experiment with Coeficient of Variation (CV) reduction (batch size $160$, $600$ epochs) required approximately 20 hours.

\subsection*{User workflow}
This section provides a high-level guide for applying our method. We outline how to define a problem, which hyperparameters matter in practice, how to configure logging, how to store trained models, and how to visualize results.

\subsubsection*{Conceptualization of the problem}~
This is the first---and one of the most critical---steps for a successful application of our method. In our setting, it starts with choosing the application domain: whether to use a deterministic or stochastic CRN interpretation, and selecting a reaction library. These choices determine which backend is used to evaluate candidate CRNs inside the training loop.

To define the task, the user must specify (i) a loss function, (ii) a set of inputs, and (iii) initial conditions. Since many applications can be formulated as transient tracking problems, we provide utilities to construct distance-from-reference loss functions among others.

As an example, consider a task that searches for circuits exhibiting RPA. We used a single initial condition (the zero state), allowed deterministic CRNs with mass-action kinetics, selected a small set of inputs and disturbances ($u_{\mathrm{in}}\in\{1,2,3\}$ and $u_{\mathrm{disturbance}}\in\{0.5,1,1.5\}$), and defined the loss as the time integral of the tracking error between a designated output species and a reference proportional to $u_{\mathrm{in}}$ as detailed in a previous section. This encourages both accurate setpoint tracking and fast responses.

\subsubsection*{Hyperparameters}~
After defining the problem, the method hyperparameters must be specified; in
practice, only a small subset has a noticeable effect on performance. A compact
reference is provided in Table~\ref{tab:hyperparams}, which reports the values
used for this task (and can serve as a reasonable default configuration).

\subsubsection*{Training and logging}~
We provide an interface for experiment tracking via \texttt{Comet.ml}, enabling
users to monitor training progress and to log each component of the objective
(e.g., reward, entropy bonuses, and auxiliary terms) separately. In addition,
we support task-dependent visualizations of candidate CRN behaviors. For
example, in the setting considered above, we render the transient trajectories
of the designated output species under multiple input levels and disturbance
instances to directly assess tracking and robustness.

At the end of training, users can export the trained RL agent together with a
snapshot of the Hall of Fame. The Hall of Fame stores the highest-performing
CRNs encountered during optimization and should be interpreted as the primary
output of the search procedure (i.e., a curated set of promising designs).

\begin{table*}[t]
\centering
\small
\begin{tabular}{p{0.22\linewidth} p{0.43\linewidth} p{0.15\linewidth} p{0.12\linewidth}}
\hline
\textbf{Hyperparameter} & \textbf{Description} & \textbf{Default value} & \textbf{Impact} \\
\hline

\multicolumn{4}{l}{\textbf{Entropy}} \\
\hline
$\lambda_H$ & Weight of the entropy bonus to promote exploration. & $5\times 10^{-3}$ & high \\
$w$ & Entropy weight encouraging diversity in proposed CRN parameters. & $1/5$ & high \\

\hline
\multicolumn{4}{l}{\textbf{Risk}} \\
\hline
$p$ & Risk level: use only the top $p$-percentile topologies to compute updates. & $0.90$ & high \\

\hline
\multicolumn{4}{l}{\textbf{Self-Imitation Learning (SIL)}} \\
\hline
$\lambda_{\mathrm{SIL}}$ & Weight of the SIL loss to reduce forgetting by reinforcing past good solutions. & $1.0$ & low \\
$n_{\mathrm{SIL\text{-}HOF}}$ & Replay buffer size (Hall-of-Fame): number of CRNs stored for SIL. & $50$ & low \\

\hline
\multicolumn{4}{l}{\textbf{Agent}} \\
\hline
$n_{\mathrm{epochs}}$ & Number of training epochs. & $300$ & moderate \\
$\mathrm{net\_width}$ & Hidden-layer width of the policy feed-forward network. & $10240$ & low \\
$\mathrm{net\_depth}$ & Number of hidden layers in the policy feed-forward network. & $5$ & low \\
$\mathrm{batch\_size}$ &  mini-batch size & 1024 & moderate \\
\hline
\end{tabular}
\caption{Hyperparameters used for the \crngen.}
\label{tab:hyperparams}
\end{table*}

\subsection*{Parameters of discovered I/O CRNs}
All parameters for the discovered I/O CRNs presented in this manuscript are provided in the Supplementary Material. We also provide an explicit rewriting of each CRN in terms of reaction equations (as a complement to the graphical representations).

\clearpage
\setcounter{extfigure}{0}
\begin{extfigure*}[ht!]
  \centering
  \includegraphics[scale=1]{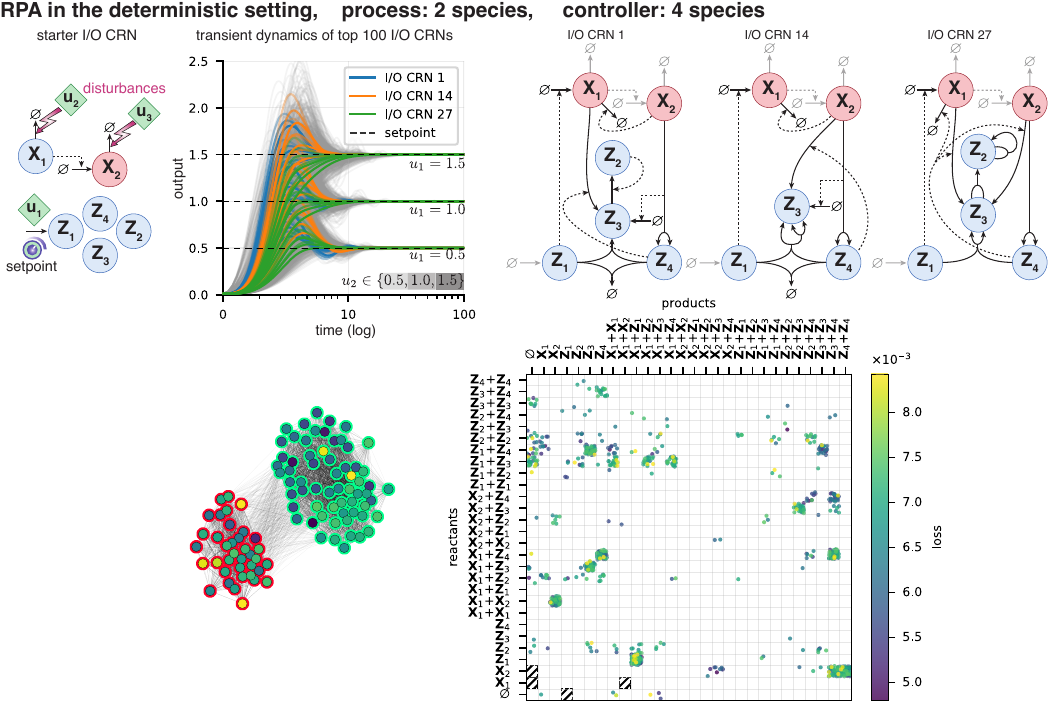}
  \caption{{\bf Deterministic RPA with 6 species.} \crngen\ scales to a higher-dimensional setting with two process species, the regulated output \species{X_2} (red) and an additional species \species{X_1}, together with four controller species \species{Z_1}--\species{Z_4} (blue). The setpoint is specified by $u_1$, while disturbances $u_2$ and $u_3$ modulate degradation of \species{X_1} and \species{X_2}, respectively. Starting from this multi-species \startercrn, \crngen\ generates candidate I/O CRNs by appending up to eight doubly bimolecular mass-action reactions and optimizing the same setpoint-tracking objective as in Fig.~\ref{Fig:RPA}A. The center plot shows output trajectories for all generated solutions (gray), with three representative networks highlighted (I/O CRNs 1, 14, and 27) and their corresponding topologies shown to the right. The graph visualization and reactant--product incidence map summarize topological diversity and reaction usage across the generated collection, as in Fig.~\ref{Fig:RPA}A.}
  \label{EDF:RPA}
\end{extfigure*}

\clearpage
\begin{extfigure*}[ht!]
  \centering
  \includegraphics[scale=1]{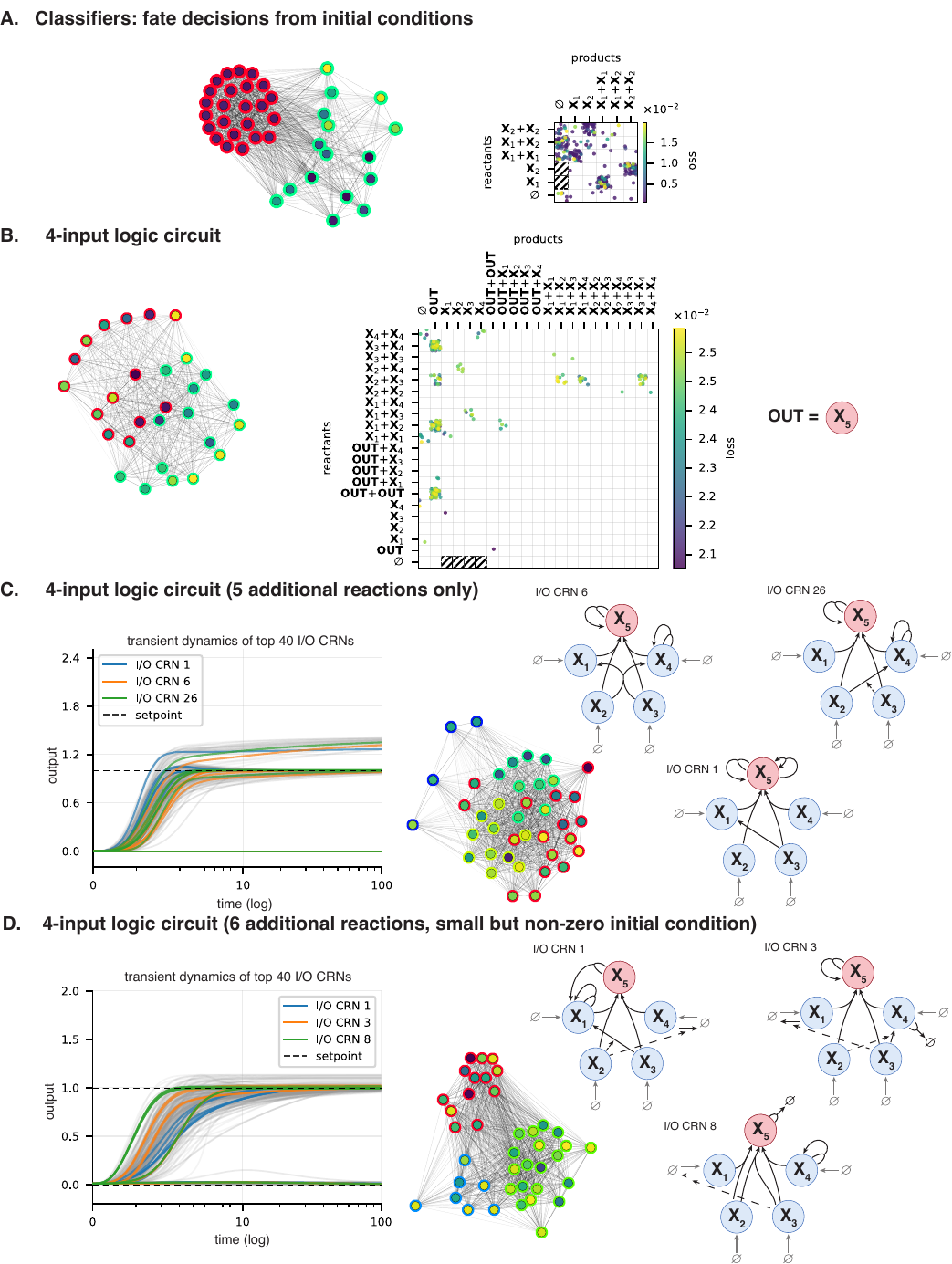}
  \caption{{\bf Classifiers and logic circuits.} The figure caption is on the next page.}
  \label{EDF:Classifier}
\end{extfigure*}

\begin{extfigure*}[ht!]\ContinuedFloat
  \centering
  \caption{{\bf Classifiers and logic circuits.} {\bf (a)} and {\bf (b)}: The graph visualization and reactant--product incidence map summarizing topological diversity and reaction usage across the generated collections of Fig.~\ref{figure:Classifier}A and B respectively.
  \textbf{(c)} Restricting the search to at most 5 appended reactions (instead of 6 as in Fig.~\ref{figure:Classifier}B), circuits that remain correct under thresholding but exhibit a more analog implementation of the logic; in particular, highlighted examples show a noticeable overshoot above the nominal ``high'' level in the all-inputs-high condition, in contrast to the cleaner saturation obtained in Fig.~\ref{figure:Classifier}B when allowing one additional reaction.
  \textbf{(d)} This panel is the same as Fig.~\ref{figure:Classifier}B with only one difference: the initial conditions are not zero to avoid exploiting unstable fixed points at the origin. \crngen also succeeded in generating diverse I/O CRNs under these conditions.}
\end{extfigure*}

\clearpage
\begin{extfigure*}[ht!]
  \centering
  \includegraphics[scale=1]{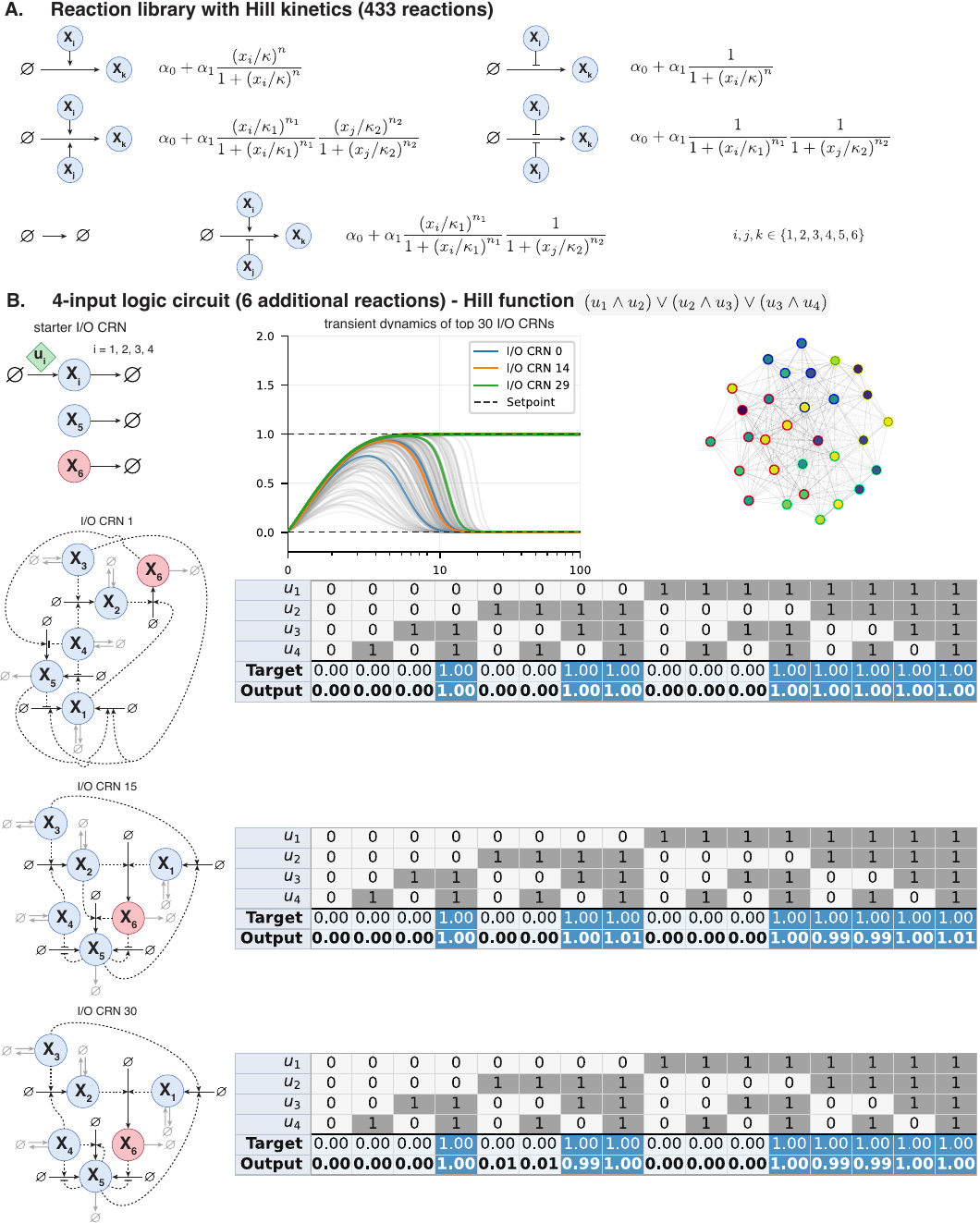}
  \caption{{\bf Logic design with Hill-type kinetics.} The figure caption is on the next page}
  \label{EDF:HF}
\end{extfigure*}

\begin{extfigure*}[ht!]\ContinuedFloat
\caption{\textbf{Logic design with Hill-type kinetics.}
\textbf{(A) Reaction library with Hill kinetics.} To test \crngen with reactions governed by multiple parameters, we construct a reaction library of 433 admissible reaction templates whose propensities follow Hill-type activation and repression forms (schematics). These templates include single-enzyme and multi-enzyme Hill functions that modulate effective production rates, yielding a larger action space for logic synthesis.
\textbf{(B) 4-input logic circuit using Hill functions.} \crngen\ starts from a \emph{starter I/O CRN} (left) containing species \species{X_1}--\species{X_6}, with four external inputs $u_i$ ($i=1,\dots,4$) that modulate the production rates of \species{X_1}--\species{X_4}. All species are also assumed to dilute at a constant rate. The objective is to implement the target 4-input Boolean function shown above the panel. \crngen\ appends up to \textbf{6 additional reactions} selected from the Hill library and optimizes a terminal-time truth-table loss identical to that in Fig.~\ref{figure:Classifier}B. The center plot shows output transients for the \textbf{top 30} topologically unique generated I/O CRNs (gray), with three representative solutions highlighted (colored; I/O CRNs 1, 15, and 30) and their corresponding topologies shown to the left. The network visualization (right) summarizes topological diversity across the 30 solutions. The truth-table panels (bottom) compare target outputs to simulated outputs across all $2^4$ input combinations for the representative networks, demonstrating accurate multi-input logic computation with Hill-type kinetics. Note that, compared to Fig.~\ref{figure:Classifier}B, this Hill library is less flexible: only production reactions can be selected, whereas degradation reactions are not available. This constraint makes the task more demanding and motivates inclusion of an additional intermediate species (\species{X_5}) to achieve high-precision logic behavior.}
\noindent\rule{\linewidth}{0.4pt}
\end{extfigure*}

\begin{extfigure*}[ht!]
  \includegraphics[scale=1]{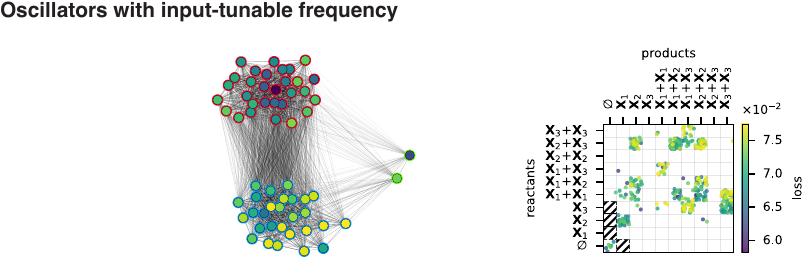}
  \caption{{\bf Oscillators with input-tunable frequency.} The graph visualization and reactant--product incidence map summarizing topological diversity and reaction usage across the generated collections of Fig.~\ref{fig:Oscillators}B. }
  \label{EDF:Oscillators}
\end{extfigure*}

\clearpage
\begin{extfigure*}[ht!]
\centering
\includegraphics[scale=1]{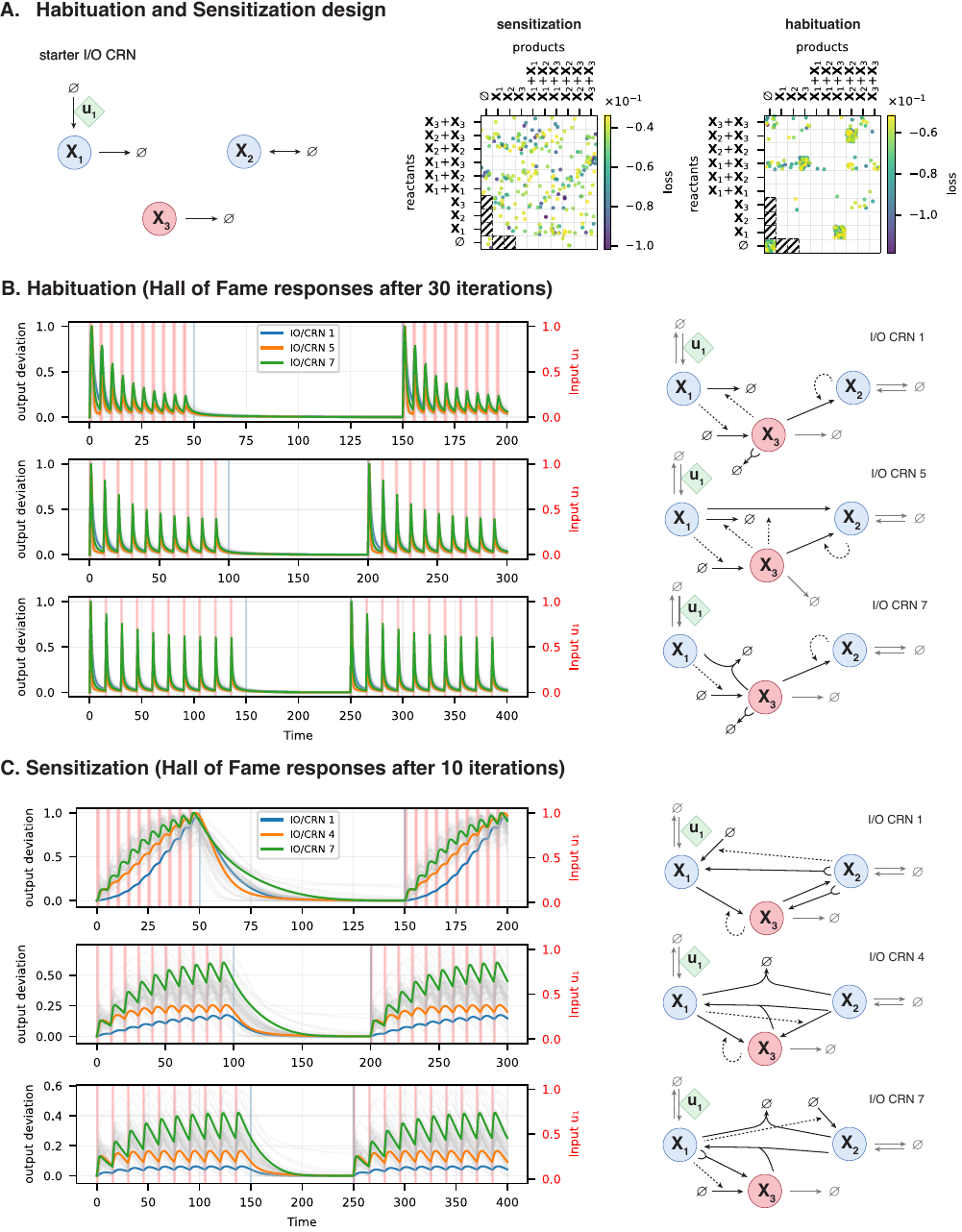}
\caption{{\bf Molecular habituation and sensitization. The figure caption is on the next page.}} 
\label{EDF:Habituation}
\end{extfigure*}

\begin{extfigure*}[ht!]\ContinuedFloat
\caption{\textbf{Design of biomolecular circuits exhibiting habituation and sensitization.}
\textbf{(a) Problem setup and reaction signatures.} \crngen\ starts from a \emph{starter I/O CRN} with three species, an external input $u_1$, and a designated output species \species{X_3} (red). The design objective is to generate I/O CRNs whose output exhibits either \emph{sensitization} (progressively increasing responses under repeated stimulation) or \emph{habituation} (progressively decreasing responses), while preserving spontaneous recovery after an OFF gap (see Methods). The reactant--product incidence maps at right summarize reaction usage for the corresponding solution sets: each point denotes a reaction, and point color indicates the loss of I/O CRNs in which that reaction appears.
\textbf{(b) Habituation.} After only 30 iterations, \crngen\ identifies a hall-of-fame set of habituating I/O CRNs. The left panels show representative output trajectories for three highlighted networks (I/O CRNs 1, 5, and 7; colored curves), with the input pulse train $u_1$ overlaid in red. Across repeated pulses, the output peaks progressively attenuate, and after the OFF gap the response recovers before adapting again during the second pulse train, consistent with habituation and spontaneous recovery. Furthermore higher frequencies exhibit faster habituation. Gray trajectories indicate the remaining hall-of-fame solutions. The corresponding I/O CRN topologies are shown at right. These topologies show negative feedback structures, incoherent feedforward loop structure as reported in \cite{smart2024minimal, eckert2024biochemically}, and buffering structures where \species{X_2} monitors the output and converts it to itself faster when the next pulse comes in.
\textbf{(b) Sensitization.} After only 10 iterations, \crngen\ likewise identifies a hall-of-fame set of sensitizing I/O CRNs. The left panels show representative trajectories for I/O CRNs 1, 4, and 7, again with the input pulse train overlaid in red. In contrast to (c), repeated stimulation now produces progressively larger output peaks, followed by recovery during the OFF gap and renewed sensitization under the second pulse train. The associated I/O CRN topologies are shown at right. Together, these results show that \crngen\ can generate distinct circuit families that realize opposite forms of adaptive pulse-train response using the same overall design framework.}
\end{extfigure*}

\end{document}